\documentclass[journal]{IEEEtran}
\usepackage{cite}

\ifCLASSINFOpdf
\else
\fi

\usepackage{amsfonts}
\usepackage{amsthm, amssymb}
\usepackage{mathrsfs}
\usepackage{amsmath}
\usepackage{booktabs}
\usepackage{multirow}
\usepackage{multicol}
\usepackage{array}
\usepackage{graphicx}
\usepackage{subfigure}
\usepackage{algorithm}
\usepackage{algorithmic}
\usepackage{stfloats}
\usepackage{titlesec}
\usepackage{indentfirst}

\titlespacing*{\section}{0pt}{\baselineskip}{0.25\baselineskip}
\titlespacing*{\subsection}{0pt}{0.25\baselineskip}{0pt}

\begin{document}
%
\title{Affine Invariant Semi-Blind Receiver: Joint Channel Estimation and High-Order Signal Detection for Multiuser Massive MIMO-OFDM Systems}

%
%
%

\author{
    Erdeng~Zhang,\ 
    Shuntian~Zheng,~\IEEEmembership{Student Member,~IEEE},
    Sheng~Wu,~\IEEEmembership{Member,~IEEE},
    Haoge~Jia,~\IEEEmembership{Member,~IEEE},
    Zhe~Ji,~\IEEEmembership{Member, IEEE},
    and Ailing Xiao,~\IEEEmembership{Member, IEEE}
\thanks{Erdeng Zhang, Shuntian Zheng, Sheng Wu, Haoge Jia, Zhe Ji and Ailing Xiao are with the School of Information and Communication Engineering, Beijing University of Posts and Telecommunications, Beijing 100876, China (email: {erdengzhang, shuntianzh, thuraya, jhg, jiz18, xiao\_ailing}@bupt.edu.cn).}
}

\maketitle

\begin{abstract}
Massive multiple input and multiple output (MIMO) systems with orthogonal frequency division multiplexing (OFDM) are foundational for downlink multi-user (MU) communication in future wireless networks, for their ability to enhance spectral efficiency and support a large number of users simultaneously. 
However, high user density intensifies severe inter-user interference (IUI) and pilot overhead. Consequently, existing blind and semi-blind channel estimation (CE) and signal detection (SD) algorithms suffer performance degradation and increased complexity, especially when further challenged by frequency-selective channels and high-order modulation demands.
To this end, this paper proposes a novel semi-blind joint channel estimation and signal detection (JCESD) method. 
Specifically, the proposed approach employs a hybrid precoding architecture to suppress IUI.
Furthermore we formulate JCESD as a non-convex constellation fitting optimization exploiting constellation affine invariance. 
Few pilots are used to achieve coarse estimation for initialization and ambiguity resolution.
For high-order modulations, a data augmentation mechanism utilizes the symmetry of quadrature amplitude modulation (QAM) constellations to increase the effective number of samples.
To address frequency-selective channels, CE accuracy is then enhanced via an iterative refinement strategy that leverages improved SD results.
Simulation results demonstrate an average throughput gain of 11\% over widely used pilot-based methods in MU scenarios, highlighting the proposed method's potential to improve spectral efficiency.

\end{abstract}


%
\IEEEpeerreviewmaketitle

\section{Introduction}

\IEEEPARstart{T}{he} explosive demand for high-speed and reliable wireless services in beyond 5G and 6G networks, such as enhanced mobile broadband, massive machine-type communications, ultra-reliable and low-latency communications, is driving the adoption of massive MIMO combined with OFDM for downlink MU transmission scenarios \cite{ref13}. 
These systems are characterized by large-scale antenna arrays, densely deployed users, and frequency-selective channels \cite{ref14}. 
To fully exploit the potential of MIMO-OFDM systems, ensuring high spectral efficiency hinges critically on accurate channel estimation and signal detection (CESD) \cite{ref16}. 
However, as the number of users increases, severe IUI lead to significant degradation in the accuracy of CESD in downlink transmissions, \cite{ref36}. \par

To maintain CESD accuracy amid growing user density, pilot-based approaches require the number of pilot symbols to scale with the number of users, leading to significant pilot overhead and reduced transmission efficiency \cite{ref37}, \cite{ref38}. 
At the same time, the intensified IUI and overlapping pilot resources, commonly referred to as pilot contamination, severely degrades the quality of estimated channel state information (CSI). 
These challenges are further exacerbated in systems employing high-order modulation like 256 quadrature amplitude modulation (QAM) and operating over frequency-selective channels \cite{ref14}.
Consequently, there is an urgent need for low-overhead, high-accuracy CESD method tailored for MU massive MIMO-OFDM systems in interference-dominant environments.\par


Current CE methods can generally be classified into three categories. Pilot-based CE, which is widely used in communication standards due to its ease of implementation and high robustness \cite{ref19}. A major drawback of this method is the substantial pilot overhead. Blind CE, which does not require any pilot symbols, obtains CSI based on the statistical properties of channels or transmitted symbols \cite{ref20}, \cite{ref21}. Although pilot-free, such methods are still affected by ambiguity issues \cite{ref10} and are associated with higher computational complexity \cite{ref22}. \par

Semi-blind CE, which mitigates ambiguity by introducing a small number of pilot symbols \cite{ref23}, \cite{ref24}. The accuracy of CE is improved while reducing pilot overhead. 
In blind and semi-blind CE methods, second-order statistics of the received signals, such as covariance and correlation, are the most commonly used statistical properties \cite{ref25}. Among these, subspace-based methods, which perform singular value decomposition (SVD) or eigenvalue decomposition (EVD) on the covariance matrix of the received signal, have been extensively applied \cite{ref23}, \cite{ref26}, \cite{ref27}. Additionally, some semi-blind CE methods leverage the correlation of received signals for implementation \cite{ref28}. However, many received symbols are required to ensure CE accuracy. Although numerous studies have proposed techniques to improve the convergence speed \cite{ref29}, \cite{ref30}, a substantial number of symbols are still needed. In particular, Dean \textit{et al.} \cite{ref1} formulate the JCESD problem as fitting a parallelepiped based on the affine invariance between the transmitter and receiver constellation and further enhance it in \cite{ref31} by utilizing linear and mixed-integer programming techniques. Elina and Bhaskar in \cite{ref39} develop an expectation-maximization (EM) algorithm for semi-blind CE. However, within massive MIMO systems, these methods still require a considerable number of received symbols. Moreover, these CE methods predominantly focus on block fading channels, while significant challenges remain in frequency-selective channels. \par

In addition to the aforementioned limitations, existing semi-blind CE methods still suffer from several limitations. In practice, the highest feasible modulation order, such as 256-QAM, is typically selected based on the signal-to-interference-plus-noise ratio (SINR) to improve spectral efficiency. However, existing works mainly focus on low-order QAM or \textit{M}-pulse amplitude modulation (PAM). For instance, the method in \cite{ref1} necessitates \textit{M}-PAM modulation, while the approaches in \cite{ref20}, \cite{ref22}, \cite{ref24} and \cite{ref27} employ 16-QAM. Additionally, the methods in \cite{ref23}, \cite{ref26}, \cite{ref28} and \cite{ref29} utilize 4-QAM or quadrature phase shift keying (QPSK), while those in \cite{ref21} and \cite{ref25} only support amplitude shift keying (ASK) or phase shift keying (PSK). Although the aforementioned methods achieve high CE accuracy, the system throughput is constrained by the reliance on the use of low-order modulation schemes. \par

Meanwhile, the aforementioned approaches mainly assume a single-user system, where only noise needs to be considered and IUI is neglected. In MU downlink scenarios, IUI is typically unknown and can significantly degrade the accuracy of blind or semi-blind CE. In addition, the dimensions of the channel matrices are determined by antenna arrays \cite{ref32}, and in massive MIMO systems, the high-dimensional channel matrix introduces new challenges for blind and semi-blind CE. To address this issue, hybrid precoding architectures are widely adopted in massive MIMO systems, employing a limited number of radio frequency (RF) chains to control a large array of antennas through the analog precoding \cite{ref33}, eliminating IUI by the digital precoding \cite{ref34}. However, the zero forcing (ZF) precoder in \cite{ref34} is applicable only when the number of RF chains is equal to the number of data streams. When the number of RF chains exceeds the number of data streams, the eigen zero forcing (EZF) precoder in \cite{ref3} cannot eliminate IUI, and further details on this issue will be given in Section \ref{sec2}. \par

In this context, a semi-blind JCESD method is proposed for downlink MU massive MIMO-OFDM systems. The primary advantage of the proposed algorithm lies in its capability to support up to 256-QAM in frequency-selective channels, achieving higher overall throughput compared to pilot-based CE methods. Specifically, the contributions of
this paper can be summarized as follows:
\begin{itemize}
    \item Built upon a hybrid analog-digital precoding architecture, this work introduces a joint transceiver digital precoding scheme to suppress IUI, enabling accurate CESD in downlink MU massive MIMO-OFDM scenarios. On this basis, a novel constellation fitting based formulation of the semi-blind JCESD problem is developed by the affine invariance between transmitter and receiver constellations. 
    \item A constraint-augmented optimization framework is designed specifically for QAM constellations. To mitigate permutation and symbol ambiguities, a small number of orthogonal pilot symbols are inserted into each data stream to provide initialization and identification, while additional algebraic constraints are imposed to reduce the degrees of freedom during the iterative solution process, improving convergence and robustness.
    \item An enhanced iterative JCESD approach for frequency-selective channels is proposed through a block-wise processing and iterative refinement strategy. To address sample insufficiency in high-order modulation, a data augmentation mechanism is introduced based on the inherent symmetry of QAM constellations, substantially increasing the number of effective samples for optimization.
    \item Extensive simulations validate that the proposed method consistently outperforms conventional pilot-based and existing semi-blind algorithms in the throughput. In particular, under practical downlink MU massive MIMO-OFDM settings,  the proposed method achieves an average throughput gain of approximately 11\% across various signal-to-noise ratio (SNR) levels.
\end{itemize}

The remainder of this paper is organized as follows.
The system model of the MU massive MIMO-OFDM systems and the channel model are described in Section \ref{sec2}. 
The proposed semi-blind JCESD algorithm are investigated in Section \ref{sec3}. 
Section \ref{sec4} evaluates the performance of the proposed algorithms. 
Finally, Section \ref{sec5} provides a conclusion of this paper.

\textit{Notation:} In the rest of this paper, we adopt the following notations: bold uppercase \(\mathbf{A}\) denotes a matrix, bold lowercase \(\mathbf{a}\) denotes a vector. Additionally, \(\left\| \mathbf{A} \right\|_2\), \({\left\| {\mathbf{A}} \right\|_\infty }\) and \({\left\| {\mathbf{A}} \right\|_{\mathsf{F}}}\) denote the \(l_2\)-norm ,infinity norm and Frobenius norm of \(\mathbf{A}\), respectively. \(\mathbf{A}^{-1}\), \(\mathbf{A}^{\mathsf{T}}\), and \(\mathbf{A}^{\mathsf{H}}\) are the represent the inverse, transpose, and Hermitian transpose of \(\mathbf{A}\). \(\mathfrak{Re}(\mathbf{A})\) and \(\mathfrak{Im}(\mathbf{A})\) are the real and imaginary parts of \(\mathbf{A}\). \(\text{det}(\mathbf{A})\) is defined as the determinant of \(\mathbf{A}\). \(\text{Blockdiag}[\mathbf{a}_1,\ldots,\mathbf{a}_n]\) denotes a block diagonal matrix with its diagonal blocks being \(\mathbf{a}_1,\ldots,\mathbf{a}_n\). \(\text{diag}(\mathbf{A})\) refers to the main diagonal of \(\mathbf{A}\). \(\mathbf{A}_{ij}\) represents the element in the \(i^\text{th}\) row and \(j^\text{th}\) column of \(\mathbf{A}\).

\section{System Model}
\label{sec2}
This section presents the system model based on hybrid precoding and a sub-connected architecture, followed by a description of the adopted channel model. \par

\subsection{Hybrid Precoding Architecture}
\begin{figure*}[htp]
    \centering
    \setlength{\abovecaptionskip}{0.cm}
    \includegraphics[width=0.8\textwidth]{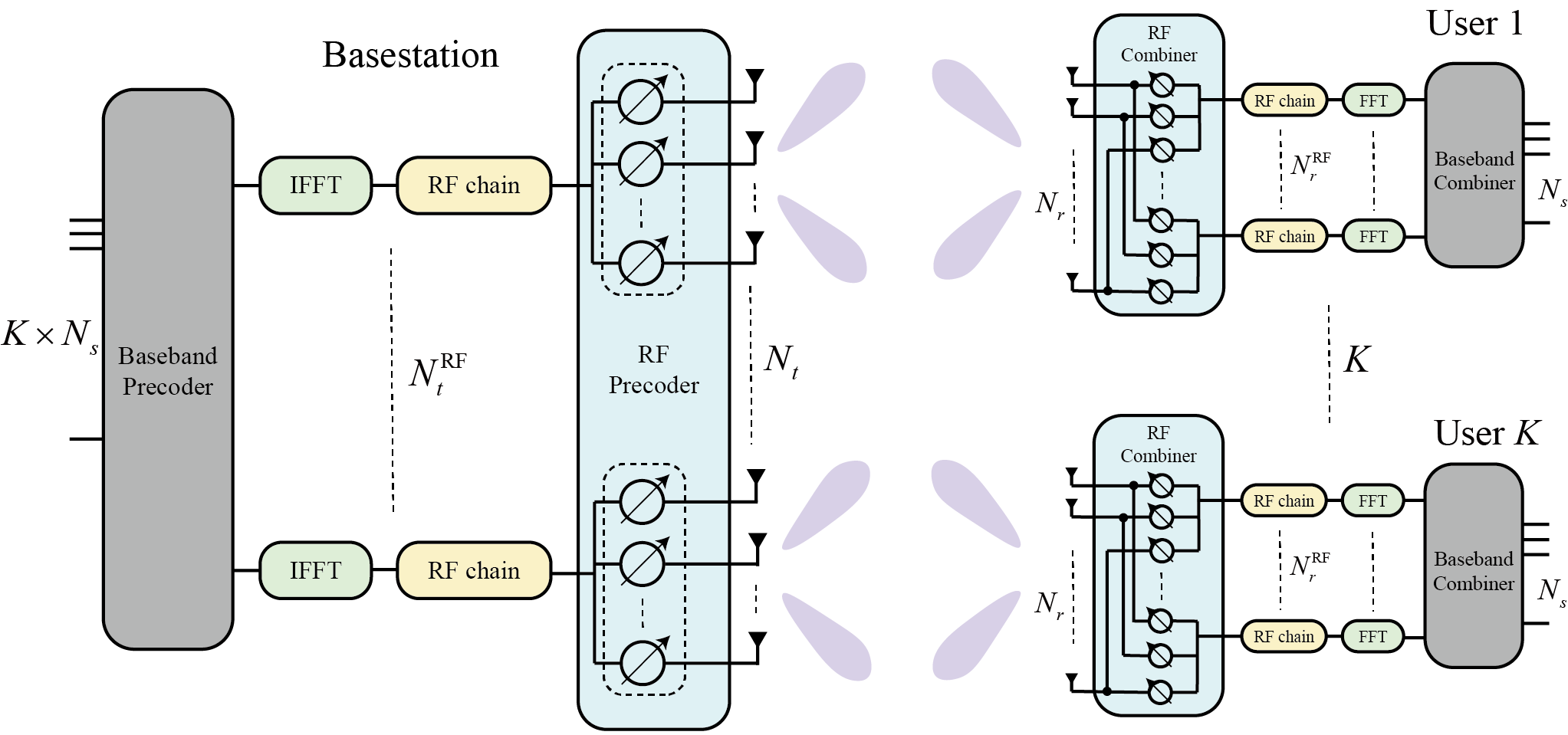}
    \caption{Structure of the hybrid precoder and combiner at the transmitter and receivers, where the RF precoder and combiner adopt a sub-connected phase shifter network.}
    \label{fig1}
    \vspace{-0.25cm}
\end{figure*}

Consider a downlink MU massive MIMO-OFDM system operating \(J\) subcarriers and \(T\) OFDM symbols, where a base station (BS) equipped with \(N_\text{t}\) antennas transmits data to \(K\) users.
Each user is equipped with \(N_\text{r}\) antennas and receives \(N_\text{s}\) data streams. The system utilizes a hybrid precoding architecture with sub-connected phase shifters as depicted in Fig. \ref{fig1}. 
The transmitter and receivers are equipped with \(N^{\text{RF}}_\text{t}\) and \(N^{\text{RF}}_\text{r}\) radio frequency (RF) chains, respectively. Each RF chain of the transmitter is connected to \({N_\text{t}}/N^{\text{RF}}_\text{t}\) antennas. 
The hybrid precoder used in the system consists of two parts: frequency-flat analog precoder \({\mathbf{F}_{\text{RF}}} \in \mathbb{C}^{N_\text{t} \times N^{\text{RF}}_\text{t}}\) and frequency-selective digital precoder \({\mathbf{F}_{\text{BB},k}}[j] \in \mathbb{C}^{N^{\text{RF}}_\text{t} \times N_\text{s}}\). 
The hybrid precoder generally takes the form \({\mathbf{F}_k}[j]={\mathbf{F}_\text{RF}{\mathbf{F}_{\text{BB},k}}[j]} \in {\mathbb{C}^{N_\text{t} \times N_\text{s}}}\) \cite{ref2}. 
After transmission through the channel, the hybrid combiner \({\mathbf{W}_k}[j]={\mathbf{W}_{\text{RF},k}{\mathbf{W}_{\text{BB},k}}}[j] \in {\mathbb{C}^{N_\text{r} \times N_\text{s}}}\) is applied to the received signal at the \(k^\text{th}\) user, to obtain the baseband received signal for a specific OFDM symbol as 
\begin{equation}
    \setlength\abovedisplayskip{3pt}
    \setlength\belowdisplayskip{3pt}
    \begin{split}
        \label{e3}
        {{\mathbf{y}}_k}[j] =& \underbrace {{\mathbf{W}}_{{\text{BB}},k}^{\mathsf{H}}[j]{\mathbf{W}}^{\mathsf{H}}_{{\text{RF,}}k}{{\mathbf{H}}_k}[j]{{\mathbf{F}}_{{\text{RF}}}}{{\mathbf{F}}_{{\text{BB,}}k}}[j]{{\mathbf{x}}_k}[j]}_{{\text{desired signal}}}\\
        &+\underbrace {\sum\limits_{m \ne k}^K {{\mathbf{W}}^{\mathsf{H}}_{{\text{BB}},k}[j]{\mathbf{W}}_{{\text{RF,}}k}^{\mathsf{H}}{{\mathbf{H}}_k}[j]{{\mathbf{F}}_{{\text{RF}}}}{{\mathbf{F}}_{{\text{BB,}}m}}[j]{{\mathbf{x}}_m}[j]} }_{{\text{IUI}}}\\ 
        &+\underbrace {{\mathbf{W}}_{{\text{BB}},k}^{\mathsf{H}}[j]{\mathbf{W}}_{{\text{RF}},k}^{\mathsf{H}}{\mathbf{n}}}_{{\text{effective noise}}},
    \end{split}
\end{equation}
where \({\mathbf{H}_k[j]}\in {\mathbb{C}^{N_\text{r}\times N_\text{t}}}\) denotes the frequency-domain channel matrix and \(\mathbf{x}_k[j] \in {\mathbb{C}^{N_\text{s}\times 1}}\) is the transmitted signal. The subscripts \(k\), and \(j\) represent the \(k^{\text{th}}\) user, and the \(j^{\text{th}}\) subcarrier. \({{\mathbf{n}}} \in {\mathbb{C}^{{N_\text{r}} \times 1}}\) is the additive white Gaussian noise (AWGN) vector and has i.i.d. entries drawn from  \(\mathcal{CN}({\mathbf{0}},\sigma^2{\mathbf{I}})\). The transmitted symbol vectors for different users are assumed to be independent. \par

The subsequent subsections elaborate on the design methodologies of analog and digital precoders and combiners, aiming to mitigate IUI.

\subsubsection{Design of Analog Precoder and Combiner}
In the sub-connected structure, the analog precoding matrix adopts a block-diagonal structure, which can be denoted as
\begin{equation}
    \label{e1}
    \setlength\abovedisplayskip{3pt}
    \setlength\belowdisplayskip{3pt}
    {{\mathbf{F}}_{\text{RF}}} = \text{Blockdiag}[{\mathbf{f}}_1,{\mathbf{f}}_2, \ldots ,{\mathbf{f}}_{N_\text{t}}/N^{\text{RF}}_\text{t}],
\end{equation}
where \({\mathbf{f}}_m\in{\mathbb{C}^{{N_\text{t}}/N^{\text{RF}}_\text{t}\times 1}}\) for \(m=1, \ldots, {N_\text{t}}/N^{\text{RF}}_\text{t}\). The analog precoder and combiner are implemented by phase shifters, and each element in \({\mathbf{f}}_m\) and \(\mathbf{W}_{\text{RF},k}\) satisfies
\begin{equation}
    \label{e2}
    \setlength\abovedisplayskip{3pt}
    \setlength\belowdisplayskip{3pt}
    f_m = \frac{1}{\sqrt{{{N_\text{t}}/N^{\text{RF}}_\text{t}}}}{\operatorname{e}^{i\theta}},\ \text{for}\ f_m \in \mathbf{f}_m,
    \end{equation}
\begin{equation}
    \label{e4}
    \setlength\abovedisplayskip{3pt}
    \setlength\belowdisplayskip{3pt}
    w_{\text{RF},k} = \frac{1}{{\sqrt{{N_\text{r}}}}}{\operatorname{e}^{i\theta}},\ \text{for}\ w_{\text{RF},k} \in \mathbf{W}_{\text{RF},k},
\end{equation}
where \(\theta \in \left[ {0,2\pi } \right)\) refers to the phase shift value. We assume that \(\mathbf{W}_{\text{RF},k}\) is the same for all users.

\subsubsection{Design of Digital Precoder and Combiner}
The purpose of digital precoding is to diagonalize the channel matrix, thereby mitigating IUI. 
Based on the classic EZF precoding 
\footnote{In practical industrial deployments, EZF precoding is often preferred due to its lower computational complexity. It is also more adaptable in scenarios where the number of RF chains and data streams do not match. For the sake of practicality and alignment with real-world systems, this paper employs EZF precoding.}
in \cite{ref3}, we design a specific digital combiner and propose a joint transceiver precoding method that incorporates the characteristics of the digital combiners during the precoding process. \par
Denote \(\mathbf{\widetilde{H}}_k[j]={\mathbf{W}}_{{\text{RF,}}k}^{\mathsf{H}}{{\mathbf{H}}_k}[j]{{\mathbf{F}}_{{\text{RF}}}} \in \mathbb{C}^{N_\text{r}^{\text{RF}}\times N_\text{t}^{\text{RF}}}\) as the equivalent channel matrix, (\ref{e3}) can be simplified as
\begin{equation}
    \setlength\abovedisplayskip{3pt}
    \setlength\belowdisplayskip{3pt}
    \begin{split}
        \label{e5}
        {{\mathbf{y}}_k}[j] =& {\mathbf{W}}_{{\text{BB}},k}^{\mathsf{H}}[j]\mathbf{\widetilde{H}}_k[j]{{\mathbf{F}}_{{\text{BB,}}k}}[j]{{\mathbf{x}}_k}[j]\\
        &+\sum\limits_{m\ne k}^K {{\mathbf{W}}_{{\text{BB,}}k}^{\mathsf{H}}[j]\mathbf{\widetilde{H}}_k[j]{{\mathbf{F}}_{{\text{BB,}}m}}[j]{{\mathbf{x}}_m}[j]}\\
        &+{{\mathbf{W}}_{{\text{BB}},k}^{\mathsf{H}}[j]{\mathbf{W}}_{{\text{RF}},k}^{\mathsf{H}}{\mathbf{n}}}.
    \end{split}
\end{equation}
\par
The classic EZF precoding method first performs SVD on \(\mathbf{\widetilde{H}}_k[j]\), extracting the right singular vectors to implement zero-forcing (ZF) \cite{ref3} as
\begin{equation}
    \label{e6}
    \setlength\abovedisplayskip{3pt}
    \setlength\belowdisplayskip{3pt}
    \mathbf{\widetilde{H}}_k[j]=\mathbf{U}_k[j]\mathbf{\Sigma}_k[j]\mathbf{V}_k^
    \mathsf{H}[j],
\end{equation}
where \(\mathbf{\Sigma}_k[j]\in\mathbb{C}^{N_\text{r}^\text{RF}\times N_\text{t}^\text{RF}}\) is a diagonal matrix with its diagonal elements being the singular value of \(\mathbf{\widetilde{H}}_k[j]\) sorted in descending order.
\(\mathbf{U}_k[j]\in\mathbb{C}^{N_\text{r}^\text{RF}\times N_\text{r}^\text{RF}}\) is a unitary matrix whose columns correspond to the left singular vectors of \(\mathbf{\widetilde{H}}_k[j]\), while \(\mathbf{V}_k[j]\in\mathbb{C}^{N_\text{t}^\text{RF}\times N_\text{t}^\text{RF}}\) is composed of the right singular vectors of \(\mathbf{\widetilde{H}}_k[j]\), similarly. 
To transmit \(N_\text{s}\) data streams to the \(k^\text{th}\) user, \(N_\text{s}\) right singular vectors corresponding to the \(N_\text{s}\) largest singular values of \(\mathbf{\widetilde{H}}_k[j]\) is selected and denoted as \({{\mathbf{\widetilde V}}_k[j]} \in {\mathbb{C}^{N_\text{t}^{{\text{RF}}} \times {N_\text{s}}}}\). 
By assembling \({{\mathbf{\widetilde V}}_k[j]}\) for all users, we obtain \({\mathbf{V}}[j] \triangleq [{{{{\mathbf{\widetilde V}}}_1[j]},{{{\mathbf{\widetilde V}}}_2[j]}, \ldots,{{{\mathbf{\widetilde V}}}_K[j]}}]\), which forms the equivalent channel for digital precoding \cite{ref3}.\par
After SVD stage, ZF is applied to \(\mathbf{V}[j]\) to obtain the EZF precoding matrix as
\begin{equation}
    \label{e7}
    \setlength\abovedisplayskip{3pt}
    \setlength\belowdisplayskip{3pt}
    {{\mathbf{F}}_{{\text{BB}}}}[j]=\mathbf{V}[j]\left({\mathbf{V}^\mathsf{H}[j]\mathbf{V}[j]}\right)^{-1},
\end{equation}
where \({{\mathbf{F}}_{{\text{BB}}}}[j] \triangleq \left[ {{{\mathbf{F}}_{{\text{BB,1}}}}[j],{{\mathbf{F}}_{{\text{BB,2}}}}[j], \ldots ,{{\mathbf{F}}_{{\text{BB,}}K}}[j]} \right]\).\par
As shown in (\ref{e7}), each time the transmitter performs digital precoding, it requires matrix inversion operations with a computational complexity of \(\mathcal{O}\left( {{{\left( {K{N_\text{s}}} \right)}^3}} \right)\) for \(J\) times. 
In massive MIMO systems, where the value of \(KN_\text{s}\) is typically large, performing EZF precoding on a per-subcarrier basis leads to extremely high computational complexity. 
Therefore, the EZF precoding process is simplified by applying a common precoder across all subcarriers. Specifically, the SVD is performed on the mean of the Gram matrix of \(\mathbf{\widetilde{H}}_k[j]\) to replace (\ref{e6}) as
\begin{equation}
    \label{e8}
    \setlength\abovedisplayskip{3pt}
    \setlength\belowdisplayskip{3pt}
    {{\mathbf{\bar G}}_k} = \frac{1}{J}\sum\limits_{j = 1}^J {\mathbf{\widetilde{H}}_k^{\mathsf{H}}[j]{\mathbf{\widetilde{H}}_k[j]}},
\end{equation}
\begin{equation}
    \label{e9}
    \setlength\abovedisplayskip{3pt}
    \setlength\belowdisplayskip{3pt}
    {{\mathbf{\bar G}}_k} = {{\mathbf{\bar V}}_k}{\mathbf{\bar\Sigma}_k}{\mathbf{\bar V}}_k^{\mathsf{H}},
\end{equation}
where \({{\mathbf{\bar G}}_k}\in\mathbb{C}^{N_\text{t}^\text{RF}\times N_\text{t}^\text{RF}}\) represents the average of the Gram matrix of \(\mathbf{\widetilde{H}}_k[j]\), \({{\mathbf{\bar V}}_k}\) can be approximately regarded as the right singular matrix of \(\mathbf{\widetilde{H}}_k[j]\) for \(j=1,\dots,J\). 
The right singular vectors corresponding to the \(N_\text{s}\) largest singular values of \({{\mathbf{\bar G}}_k}\) for all users are assembled to form an equivalent channel matrix \(\mathbf{\bar V}\) similarly. Then we have a frequency-flat digital precoder as
\begin{equation}
    \label{e10}
    \setlength\abovedisplayskip{3pt}
    \setlength\belowdisplayskip{3pt}
    {\mathbf{F}}_{{\text{BB}}}=\mathbf{\bar V}\left({\mathbf{\bar V}^\mathsf{H}\mathbf{\bar V}}\right)^{-1},
\end{equation}
where \({{\mathbf{F}}_{{\text{BB}}}} \triangleq \left[ {{{\mathbf{F}}_{{\text{BB,1}}}},{{\mathbf{F}}_{{\text{BB,2}}}}, \ldots ,{{\mathbf{F}}_{{\text{BB,}}K}}} \right]\).\par
The classical EZF precoding selects the singular vectors corresponding to the \(N_\text{s}\) largest singular values to construct an equivalent channel and apply ZF, while \(\mathbf{\widetilde{H}}_k[j]\) has \(N_\text{r}^{\text{RF}}\) non-zero singular values. 
In the scenario where \(N_\text{r}^{\text{RF}}=N_\text{s}\), the method in \cite{ref3} applies ZF to the entire channel, which can eliminate most of the IUI. 
However, in massive MIMO systems, \(N_\text{r}^{\text{RF}}\) is typically much larger than \(N_\text{s}\), and therefore, the classical EZF precoding method can only mitigate partial IUI. 
To address this limitation, we design a specific digital combiner and propose a joint transceiver precoding scheme.
Specifically, the digital combiner is designed in the form of a DFT matrix as
\begin{equation}
    \label{e11}
    \setlength\abovedisplayskip{3pt}
    \setlength\belowdisplayskip{3pt}
    {{\mathbf{W}}_{{\text{BB}}}} = \frac{1}{{\sqrt {N_\text{r}^{{\text{RF}}} \times {N_\text{s}}} }}\exp \{i{\mathbf{\Theta }}\},
\end{equation}
where \(\mathbf{\Theta}\in\mathbb{R}^{N_\text{r}^{\text{RF}} \times {N_\text{s}}}\), and for each \({\theta _i} \in \mathbf{\Theta}\), \({\theta _i} \sim {\text{U}}(0,2\pi )\), i.i.d. 
It is worth noting that \({\mathbf{W}}_{\text{BB}}\) is frequency-flat, and all users share the same \({\mathbf{W}}_{\text{BB}}\).
The transmitter and receivers can obtain the same \({\mathbf{W}}_{\text{BB}}\) by setting a specific random seed. The design of the digital combiner described above is not unique, and the selected DFT matrix may not necessarily be the optimal solution. 
However, this is not our optimization objective, our purpose is to design a digital precoder to eliminate IUI. The transmitter performs EZF precoding based on \({\mathbf{W}}_{\text{BB}}\) and \({\mathbf{\widetilde{H}}_k[j]}\), (\ref{e8}) is modified as
\begin{equation}
    \label{e12}
    \setlength\abovedisplayskip{3pt}
    \setlength\belowdisplayskip{3pt}
    {{\mathbf{\bar G}}_k} = \frac{1}{J}\sum\limits_{j = 1}^J {{{\left( {{\mathbf{W}}_{{\text{BB}}}^{\mathsf{H}}{\mathbf{\widetilde{H}}_k}[j]} \right)}^{\mathsf{H}}}{\mathbf{W}}_{{\text{BB}}}^{\mathsf{H}}{\mathbf{\widetilde{H}}_k}[j]},
\end{equation}
and \({{\mathbf{\bar V}}_k}\) in (\ref{e9}) can be approximately regarded as the right singular matrix of \({\mathbf{W}}_{{\text{BB}}}^{\mathsf{H}}\mathbf{\widetilde{H}}_k[j]\in\mathbb{C}^{N_\text{s}\times N_\text{t}^{\text{RF}}}\).
Thus, we have \({\mathbf{\bar V}}_k\mathbf{F}_{\text{BB},k}\approx\mathbf{I}\) and \({\mathbf{\bar V}}_k\mathbf{F}_{\text{BB},i}\approx\mathbf{O}\) for \(i\neq k\). To control the total transmit power \(P\) and ensure it remains unchanged after digital precoding, a power control factor \(\lambda\) is typically used to normalize the signal power \cite{ref5} as
\begin{equation}
    \label{e13}
    \setlength\abovedisplayskip{3pt}
    \setlength\belowdisplayskip{3pt}
    \lambda = \sqrt {\frac{P}{{\text{Tr}\left( {{{\mathbf{F}}_{{\text{BB}}}}{\mathbf{F}}_{{\text{BB}}}^{\mathsf{H}}} \right)}}}.
\end{equation}
\par
Finally, the hybrid precoding matrix can be written in the form \(\mathbf{F}_k=\lambda\mathbf{F}_{\text{RF}}\mathbf{F}_{\text{BB},k}\).

\subsection{Channel Model}
The massive MIMO channels between the transmitter and receivers are assumed to be frequency-selective, with a delay tap length of \(N_\text{c}\) in the time domain. For the \(k^{\text{th}}\) user, denote \(\mathsf{H}_k[d]\in\mathbb{C}^{N_\text{r}\times N_\text{t}}\) as the \(d^{\text{th}}\) delay tap of the channel, where \(d=0,1,\ldots,N_\text{c}-1\). Assuming a geometric channel model \cite{ref4,ref6}, \(\mathsf{H}_k[d]\) is expressed as
\begin{equation}
\begin{split}
    \label{e14}
    \setlength\abovedisplayskip{3pt}
    \setlength\belowdisplayskip{3pt}
    \mathsf{H}_k[d]=\sqrt {\frac{{{N_\text{t}}{N_\text{r}}}}{{L{\rho_{\text{L}}}}}}\sum\limits_{\ell=1}^L&{{\alpha_{\ell,k}}} {p_{{\text{rc}}}}\left({d{T_s}-{\tau_{\ell,k}}}\right) \\&\times{{\mathbf{a}}_{\text{R}}}\left({\phi^\text{r}_{\ell,k},\theta^\text{r}_{\ell,k}}\right){\mathbf{a}}_{\text{T}}^{\mathsf{H}}\left({\phi^\text{t}_{\ell,k},\theta^\text{t}_{\ell,k}}\right),
\end{split}
\end{equation}
where \(\rho_{\text{L}}\) represents the pass loss between the BS and the \(k^{\text{th}}\) user, with the assumption that \(\rho_{\text{L}}\) is identical for all users. 
\(L\) denotes the number of paths; \(T_\text{s}\) is the sampling period; \(p_{\text{rc}}\left(\tau\right)\) represents a filter that captures the effects of pulse shaping and other low-pass filtering evaluated at \(\tau\); \(\alpha_{\ell,k}\in\mathbb{C}\) is the complex gain of the \(\ell^{\text{th}}\) path; \(\tau_{\ell, k}\) is the delay of the \(\ell^{\text{th}}\) path. Both the BS and users are assumed to use Uniform Planar Arrays (UPAs), with \({\mathbf{a}}_{\text{R}}\left({\phi^\text{r}_{\ell,k},\theta^\text{r}_{\ell,k}}\right)\) and \({\mathbf{a}}_{\text{T}}^{\mathsf{H}}\left({\phi^\text{t}_{\ell,k},\theta^\text{t}_{\ell,k}}\right)\) representing the steering vectors of the array for the receive and transmit antennas, respectively. \(\phi_{\ell,k}\in\left[0,2\pi\right]\) and \(\theta_{\ell,k}\in\left[0,2\pi\right]\) are the elevation and azimuth angle of arrival or departure of the \(\ell^{\text{th}}\) path.
The channel at the \(j^\text{th}\) subcarrier can be expressed as a sum of the different delay taps,
\begin{equation}
    \label{e15}
    \setlength\abovedisplayskip{3pt}
    \setlength\belowdisplayskip{3pt}
    \mathbf{H}_k[j]=\sum\limits_{d = 0}^{{N_\text{c}} - 1} {{\mathsf{H}_k}[d]{e^{ - i\frac{{2\pi j}}{J}d}}}.
\end{equation}
\par
Denote \(\mathbf{H}^{\text{eq}}_k[j]=\mathbf{W}_\text{BB}^{\text{H}}\mathbf{W}_{\text{RF},k}^{\text{H}}\mathbf{H}_k[j]\mathbf{F}_{\text{RF}}\mathbf{F}_{\text{BB},k}\in\mathbb{C}^{N_\text{s}\times N_\text{s}}\) as the equivalent channel, the received signal in (\ref{e3}) can be simplified as
\begin{equation}
    \label{e16}
    \setlength\abovedisplayskip{3pt}
    \setlength\belowdisplayskip{3pt}
    \mathbf{y}_k[j]=\mathbf{H}^{\text{eq}}_k[j]\mathbf{x}_k[j]+\mathbf{\widetilde n}_k[j],
\end{equation}
where \(\mathbf{\widetilde n}_k[j]\in\mathbb{C}^{N_\text{s}\times 1}\) includes both IUI and AWGN. 
Our goal is to jointly estimate the equivalent channel \(\mathbf{H}^{\text{eq}}_k[j]\) and the transmitted signal \(\mathbf{x}_k[j]\) from the received signal \(\mathbf{y}_k[j]\) using a small number of pilot symbols.

\section{Proposed Semi-Blind JCESD Algorithm}
\label{sec3}
This section introduces several enhancements to adapt the method in \cite{ref1} to high-order modulation schemes, such as 256-QAM, and complex-valued frequency-selective massive MIMO channels. 

\subsection{Semi-Blind CE Based on the Affine Invariance of Constellation}
By exploiting the affine invariance between the transmit and receive constellations, the blind JCESD problem is reformulated as a constellation fitting problem in \cite{ref1} . It focuses on a \(n\times n\) real-valued MIMO channel with block fading and AWGN, the input-output relation is characterized as
\begin{equation}
\label{e17}
    \setlength\abovedisplayskip{3pt}
    \setlength\belowdisplayskip{3pt}
    {\mathbf{y}} = {\mathbf{Hx}} + {\mathbf{n}},
\end{equation}
where \({\mathbf{H}} \in {\mathbb{R}^{n \times n}}\) and \({\mathbf{x}}\) is drawn from a standard \textit{M}-PAM constellation. Given a set of samples of \({{\mathbf{y}}_1}, \ldots {{\mathbf{y}}_m}\), and \({\mathbf{Y}} = \left[ {{{\mathbf{y}}_1}, \ldots ,{{\mathbf{y}}_m}} \right]\), consider the program
\begin{equation}
\label{e18}
    \setlength\abovedisplayskip{3pt}
    \setlength\belowdisplayskip{3pt}
    \mathop {{\text{maximize}}}\limits_{\mathbf{U}} {\text{ log }}\left| {{\text{det}}({\mathbf{U}}}) \right|
\end{equation}
\begin{equation}
\label{e19}
    \setlength\abovedisplayskip{3pt}
    \setlength\belowdisplayskip{3pt}
    {\text{subject to }}{\left\| {{\mathbf{U}}{{\mathbf{y}}_i}} \right\|_\infty } \leqslant M + c \times \sigma, \, i = 1, \ldots ,m.
\end{equation}
\par
The domain of \({\mathbf{U}}\) is all \(n\times n\) invertible matrices, and the optimal solution is given by \({\mathbf{U}} = {\mathbf{T}}{{\mathbf{H}}^{-1}}\), where \({\mathbf{T}}\) is the product of a permutation matrix and a diagonal matrix with entries \(\pm1\), referred to as an admissible transform matrix (ATM) \cite{ref9}.
The notation \(\lfloor \cdot \rceil\) denotes rounding to the nearest constellation point, \(\kappa(\cdot)\) indicates the condition number of a matrix, \(c\) is a positive constant, \(\sigma^2\) is noise variance, and \(c \times \sigma\) refers to a margin for the presence of AWGN. 
The objective of the above program is to determine a transformation matrix \(\mathbf{U}\) such that all received signal samples are mapped as closely as possible to the nearest standard constellation points. \par

The algorithm mentioned above has several limitations. 
It is specifically designed for \textit{M}-PAM, with the constraint term in (\ref{e19}) tailored to that constellation, thereby rendering it unsuitable for \textit{M}-QAM. 
Moreover, the constraint term in (\ref{e19}) applies only to the boundary constellation points, which becomes increasingly ineffective for high-order QAM constellations, where the symbol space is layered and the boundary points constitute a diminishing fraction.
In addition, the algorithm suffers from inherent ambiguity due to the invariance of the constellation under flipping and permutation operations, which prevents the differentiation between \((\mathbf{H},\mathbf{X})\) and \((\mathbf{HT},\mathbf{T}^{-1}\mathbf{X})\) in the absence of prior information.



Leveraging the structural characteristics of the \textit{M}-QAM constellation and the channel properties, the constraint term in (\ref{e19}) is revised, and additional constraints are introduced to enhance the estimation accuracy. 
To resolve the symbol ambiguity and permutation ambiguity issues in \cite{ref1}, \(N_\text{s}\) pilot symbols are introduced into each data stream. 
These pilot symbols serve to provide a more suitable initialization for the optimization algorithm, thereby enhancing its convergence speed.

\subsubsection{Data Normalization}
Since the average power of \textit{M}-QAM symbols is normalized to  1, each data stream is scaled before semi-blind CE to better match the power level, thereby enhancing convergence speed. Denote \(\mathbf{y}_{k,s}\in\mathbb{C}^m\) as the \(s^{\text{th}}\) received data streams where \(s=1,\dots, N_\text{s}\) and \(m=J\times T\) refer to the number of symbols contained in a data stream, with \(T\) is the number of OFDM symbols. Among \(\mathbf{y}_{k,s}\), let its normalized magnitude be \(a\), the normalized data stream \(\mathbf{y}_{k,s}^{\text{norm}}\) is given by
\begin{equation}
    \label{e20}
    \setlength\abovedisplayskip{3pt}
    \setlength\belowdisplayskip{3pt}
    \mathbf{y}_{k,s}^{\text{norm}}= \frac{a}{\max \left\{\mathbf{y}_{k,s}\right\}}\mathbf{y}_{k,s}.
\end{equation}
\par
After the data normalization, the  matrix composed of all received symbols on a single subcarrier is denoted as \(\mathbf{Y}_k[j]\)

\subsubsection{Design of the Constraint Term}
As mentioned above, the constraint term in (\ref{e19}) refers to the boundary of the constellation. 
The \(\mathbf{x}\) in (\ref{e17}) is drawn from \textit{M}-PAM constellation. For \textit{M}-QAM, the boundary of the constellation is represented by \(\lambda_M\). 
The \(c\times\sigma\) in (\ref{e19}) is related to noise and IUI. 
The signal-to-interference-plus-noise ratio (SINR) is defined as
\begin{equation}
    \label{e23}
    \setlength\abovedisplayskip{3pt}
    \setlength\belowdisplayskip{3pt}
    \text{SINR}_k=\frac{{\left\| {\mathbf{X}}_k \right\|_{\mathsf{F}}^2}}{\left\| {{\mathbf{\hat{X}}}_k-{\mathbf{X}}_k} \right\|_{\mathsf{F}}^2},
\end{equation}
\par
Since the power of the interference signal is unknown, alternative methods are required to estimate the SINR. 
As mentioned above, \(N_\text{s}\) pilot symbols are embedded in each data stream, and the SINR is estimated based on these pilots. 
Denote \(\mathbf{P}_{k,\text{r}}\in\mathbb{C}^{N_\text{s}\times N_\text{s}}\) as the received pilot symbols, and the corresponding transmitted pilot symbols are denoted as \(\mathbf{P}_{k,\text{t}}\). 
An initial estimate of the channel, potentially with significant errors, is obtained by least squares (LS) estimation,
\begin{equation}
    \label{e42}
    \setlength\abovedisplayskip{3pt}
    \setlength\belowdisplayskip{3pt}
    \hat{\mathbf{H}}_k^{\text{eq}}={\left({{\mathbf{P}}_{k{\text{,t}}}^{\mathsf{H}}{{\mathbf{P}}_{k{\text{,t}}}}} \right)^{-1}}{\mathbf{P}}_{k{\text{,t}}}^{\mathsf{H}}{{\mathbf{P}}_{k{\text{,r}}}}.
\end{equation}
\par
The \(\text{SINR}_k\) is then approximated as
\begin{equation}
    \label{e43}
    \setlength\abovedisplayskip{3pt}
    \setlength\belowdisplayskip{3pt}
    \text{SINR}_k \approx \frac{{\left\| {\mathbf{P}}_{k{\text{,t}}}\right\|_{\mathsf{F}}^2}}{{\left\| {{{\left( {{\mathbf{\hat H}}_k^{{\text{eq}}}} \right)}^{-1}}{\mathbf{P}}_{k{\text{,r}}} - {{\mathbf{P}}_{k{\text{,t}}}}} \right\|_{\mathsf{F}}^2}}.
\end{equation}
\par
Since the average power of \textit{M}-QAM symbols is 1, we design the constraint term in (\ref{e19}) as
\begin{equation}
    \label{e30}
    \setlength\abovedisplayskip{3pt}
    \setlength\belowdisplayskip{3pt}
    {\left\| {\mathbf{U}_k{{\mathbf{Y}}_k}} \right\|_\infty } \leqslant \lambda_M + \sqrt {\frac{1}{\text{SINR}_k}}.
\end{equation}
\par
In addition to enabling SINR estimation, the LS based CE in (\ref{e42}) also provides a reasonable initial point \(\mathbf{U}_0\) for solving the optimization problem. A theoretical analysis of its effectiveness is provided in Appendix \ref{appendixA}.

\subsubsection{Reducing the Degrees of Freedom in the Iterations}
Considering \(\mathbf{H}^{\text{eq}}_k\) is a complex-valued matrix, the optimization program, however, can only handle real-valued matrices. The corresponding real-valued matrix \({\mathbf{\widetilde H}}_k^{\text{eq}}\in {\mathbb{R}^{2{N_\text{s}}\times 2{N_\text{s}}}}\) can be expressed as \cite{ref1}
\begin{equation}
    \label{e31}
    \setlength\abovedisplayskip{3pt}
    \setlength\belowdisplayskip{3pt}
    \renewcommand{\arraystretch}{1.5} 
    {\mathbf{\widetilde H}}_k^{\text{eq}}=\left[{\begin{array}{*{20}{c}}
    {\mathfrak{Re}\left({\mathbf{H}_k^\text{eq}}\right)}&{-\mathfrak{Im}\left({\mathbf{H}_k^\text{eq}}\right)}\\
    {\mathfrak{Im}\left( {\mathbf{H}_k^\text{eq}}\right)}&{\mathfrak{Re}\left({\mathbf{H}_k^\text{eq}} \right)}\end{array}}\right],
\end{equation}
similarly, \(\mathbf{U}_k\) is mapped to a real-valued matrix \(\widetilde{\mathbf{U}}_k\). When the number of symbols is sufficiently large, \(\widetilde{\mathbf{U}}_k\) converges to \(\left({\mathbf{\widetilde H}}_k^{\text{eq}}\right)^{-1}\). 
Based on the block structure of \({\mathbf{\widetilde H}}_k^{\text{eq}}\), the block form of \(\widetilde{\mathbf{U}}_k^{-1}\) is
\begin{equation}
    \label{e32}
    \setlength\abovedisplayskip{3pt}
    \setlength\belowdisplayskip{3pt}
    \renewcommand{\arraystretch}{2.0} 
    \widetilde{\mathbf{U}}_k^{-1}=\left[{\begin{array}{*{20}{c}}
    {\mathbf{U}_{k,11}^{-1}}&{\mathbf{U}_{k,12}^{-1}}\\
    {\mathbf{U}_{k,21}^{-1}}&{\mathbf{U}_{k,22}^{-1}}\end{array}}\right],
\end{equation}
the estimation of the equivalent channel \({\mathbf{\hat H}}_k^{{\text{eq}}}\) is
\begin{equation}
    \label{e33}
    \setlength\abovedisplayskip{3pt}
    \setlength\belowdisplayskip{3pt}
    {\mathbf{\hat H}}_k^{{\text{eq}}} = \frac{\mathbf{U}_{k,11}^{-1} + {{\mathbf{U}}_{k,22}^{-1}}}{2} + \frac{{{\mathbf{U}}_{k,21}^{ - 1} - {{\mathbf{U}}_{k,12}^{ - 1}}}}{2}j.
\end{equation}
\par
A limitation of the above approach is the high degree of freedom involved in the iterations when solving the optimization problem, which results in high computational complexity. 
To address this, we introduce the following constraints in the optimization problem,
\begin{equation}
    \label{e34}
    \setlength\abovedisplayskip{3pt}
    \setlength\belowdisplayskip{3pt}
    \mathbf{U}_{k,11}^{-1}=\mathbf{U}_{k,22}^{-1}, \ \
    \mathbf{U}_{k,12}^{-1}=-\mathbf{U}_{k,21}^{-1}.
\end{equation}
\par
This constraint involves submatrices of the inverse matrix, complicating direct solution.
Considering that \(\mathbf{U}_k^{(i)}\mathbf{H}_k^{\text{eq}}=\mathbf{I}\) we have
\begin{equation}
    \label{e35}
    \setlength\abovedisplayskip{3pt}
    \setlength\belowdisplayskip{3pt}
    \begin{split}
        &\mathfrak{Re}\left(\mathbf{H}_k^{\text{eq}}\right) \mathfrak{Re}\left(\mathbf{U}_k\right)-\mathfrak{Im}\left(\mathbf{H}_k^{\text{eq}}\right) \mathfrak{Im}\left(\mathbf{U}_k\right)=\mathbf{I},\\
        &\mathfrak{Re}\left(\mathbf{H}_k^{\text{eq}}\right) \mathfrak{Im}\left(\mathbf{U}_k\right)+\mathfrak{Im}\left(\mathbf{H}_k^{\text{eq}}\right) \mathfrak{Re}\left(\mathbf{U}_k\right)=\mathbf{O},
    \end{split}
\end{equation}
thus, \(\left({\mathbf{\widetilde H}}_k^{\text{eq}}\right)^{-1}=\widetilde{\mathbf{U}}_k\). The constraint in (\ref{e34}) can be replaced by
\begin{equation}
    \setlength\abovedisplayskip{3pt}
    \setlength\belowdisplayskip{3pt}
    \label{e36}
    {\mathbf{U}_{k,11}}={\mathbf{U}_{k,22}},\,\,
    {\mathbf{U}_{k,12}}=-{\mathbf{U}_{k,21}}.
\end{equation}
\par
This reduces the degrees of freedom in the iteration to half of the original.

\subsubsection{Resolving the Ambiguity}
The optimization program defined by (\ref{e18}), (\ref{e30}) and (\ref{e36}) is solved using the Sequential Least Squares Programming (SLSQP) algorithm in \cite{ref7} and the solver in \cite{ref8} is applied to address non-linear conic optimization problems.
In the presence of the ATM \(\mathbf{T}\), relative to \(\mathbf{H}_k^{\text{eq}}\) and  \(\mathbf{X}_k\), certain columns of \(\mathbf{\hat H}_k^{\text{eq}}\) or rows of \(\mathbf{\hat{X}}_k\) may have their order permuted, this phenomenon is referred to as permutation ambiguity. 
And the elements in some columns of \(\mathbf{\hat H}_k^{\text{eq}}\) or rows of \(\mathbf{\hat{X}}_k\) may be multiplied by a constant factor of \(\pm j\), this phenomenon is called symbol ambiguity.
In the absence of other prior information, processing only the received symbols cannot resolve the ambiguity \cite{ref10}.
\par
\begin{figure*}[htp]
    \subfigure[Resolving permutation ambiguity]{
        \centering
        \includegraphics[width=0.48\textwidth]{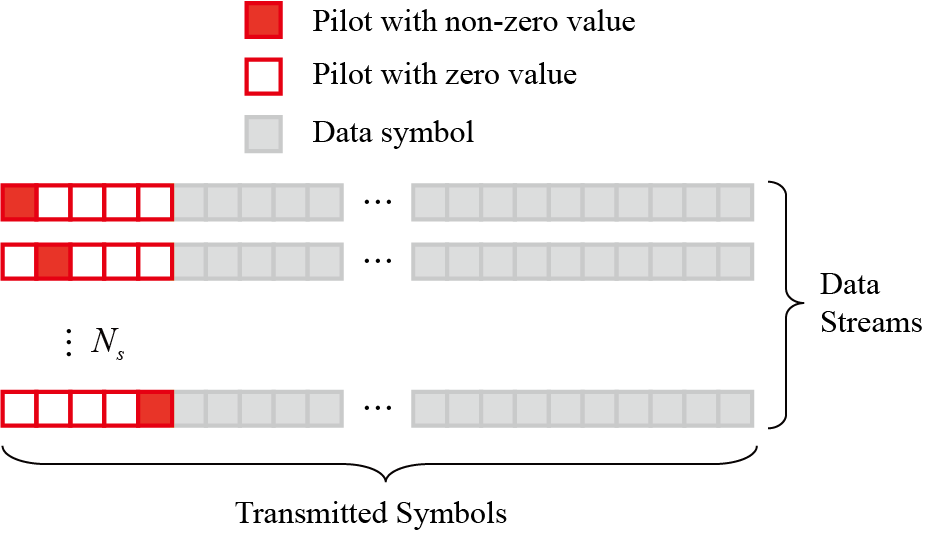}
    }
    \subfigure[Resolving symbol ambiguity]{
        \centering
        \includegraphics[width=0.48\textwidth]{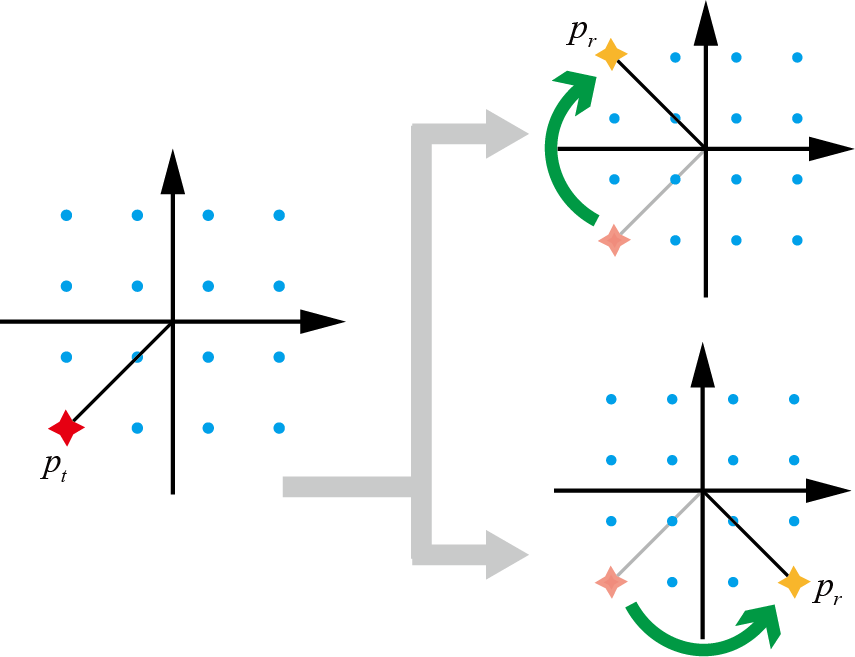}
    }
    \centering
    \setlength{\abovecaptionskip}{0.cm}
    \caption{The method of eliminating ambiguity using few pilots. Orthogonal pilots are employed between different data streams of each user. Each data stream consists of a non-zero pilot \(p_\text{t}\) and \(N_\text{s}-1\) pilots with a zero-value. The sequence of different data streams is determined based on the position of \(p_\text{t}\). Considering the effects of noise and IUI, if the phase angle variation between \(p_\text{r}\) and \(p_\text{t}\) exceeds \(45^{\circ}\), meaning that \(p_\text{r}\) and \(p_\text{t}\) are located in different quadrants of the constellation, symbol ambiguity is deemed to have occurred.} 
    \label{fig4}
\end{figure*}

As illustrated in Fig. \ref{fig4}, to resolve ambiguity, we introduce \(N_s\) pilot symbols in each data stream, with one of these pilots used to eliminate symbol ambiguity. 
Specifically, the pilot symbol \(p_\text{t}\) to be transmitted is fixed as the symbol at the bottom-left corner of the constellation. 
Symbol ambiguity can be detected by observing the angular variation between the detected pilot \(p_\text{r}\) and \(p_\text{t}\). 
These \(N_\text{s}\) pilots assign a unique identifier to each data stream, thereby enabling for the elimination of permutation ambiguity.

\par
In summary, the proposed semi-blind JCESD procedure is outlined in Algorithm \ref{alg1}. \par

\begin{algorithm}[h]
    \caption{Semi-Blind JCESD Algorithm for MIMO Channels}
    \label{alg1}
    \renewcommand{\algorithmicrequire}{\textbf{Input:}}
    \renewcommand{\algorithmicensure}{\textbf{Output:}}
    \begin{algorithmic}[1]
        \REQUIRE Received signals \({\mathbf{Y}_k}[j]\).
        \ENSURE An estimate of the channel matrix \({\hat{\mathbf{H}}_k^{\text{eq}}[j]}\), and an estimate of the transmitted symbols \({\mathbf{\hat X}_k[j]}\).
        \STATE \(\mathbf{Y}_k \triangleq \left[ {{{\mathbf{Y}}_k}[0],{{\mathbf{Y}}_k}[1], \ldots ,{{\mathbf{Y}}_k}[J]} \right]\)
        \STATE Normalize \(\mathbf{Y}_k\) via (\ref{e20})
        \IF{\(\kappa({\mathbf{Y}_k}) > \kappa_\text{max}\)}
            \RETURN FAIL
        \ENDIF
        \STATE \(\mathbf{U}_0={\left( {{\mathbf{\hat H}}_k^{{\text{eq}}}} \right)^{ - 1}}\) via (\ref{e42})
        \WHILE{\({\left\| {{{\mathbf{U}}_0}{\mathbf{Y}}_k} \right\|_\infty } > 1\)}
            \STATE \(\mathbf{U}_0=\mathbf{U}_0/2\)
        \ENDWHILE
        \STATE Perform SLSQP over (\ref{e18}), (\ref{e30}) and (\ref{e36}) starting at \(\mathbf{U}_0\) to find an optimal value \(\mathbf{U}_k\)
        \STATE \(\mathbf{\hat{H}}_k^{\text{eq}}[0:J]=\mathbf{U}_k^{-1}\)
        \STATE \(\mathbf{\hat{X}}_k[j]=\lfloor \mathbf{U}_k\mathbf{Y}_k[j]\rceil\)
        \STATE Eliminate the ambiguity
        \RETURN \({\hat{\mathbf{H}}_k^{\text{eq}}[j]}\), \({\mathbf{\hat X}_k[j]}\).
    \end{algorithmic}
\end{algorithm}

\subsection{Modified Semi-Blind JCESD for Frequency-Selective Channels}

The algorithm developed in the previous subsection assumes a block fading channel model and does not explicitly consider the frequency-selective characteristics of practical wideband channels. In contrast, this subsection presents an enhanced semi-blind JCESD framework tailored for frequency-selective fading environments, aiming to improve estimation accuracy under severe IUI and significant subcarrier variation.
\par
The semi-blind JCESD method proposed in the previous subsection processes the received data by aggregating all symbols received within a transmission time interval (TTI) into a single matrix \(\mathbf{Y}_k\), and employs a unified estimated channel matrix across all subcarriers. However, in the presence of severe frequency selectivity and severe IUI, this processing may result in significant CE errors. Therefore, the proposed method is modified to better accommodate frequency-selective channels and mitigate more pronounced IUI.

\subsubsection{Data Segmentation and Augmentation}

\begin{figure}[h]
    \centering
    \includegraphics{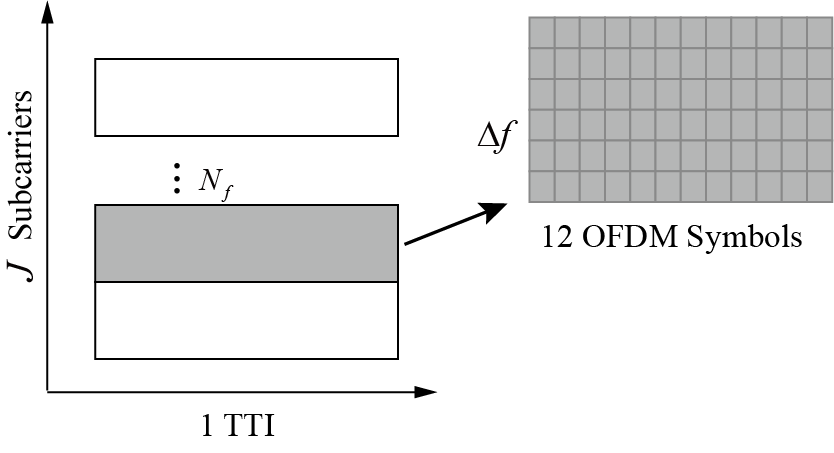}
    \caption{The division of the received symbols from the whole TTI into multiple blocks.}
    \label{fig3}
\end{figure}

The entire TTI is divided into \(N_\text{f}\) blocks, each block spanning \(\Delta f=J/N_\text{f}\) subcarriers, as shown in Fig. \ref{fig3}. The semi-blind CE is then performed for each block separately to obtain different CE results.
The \(i^{\text{th}}\) block contains \(m'=\Delta f\times T\) received symbols, where \(i=1,\dots,N_\text{f}\), these received symbols form the matrix \(\mathbf{Y}_{k,i}\in\mathbb{C}^{N_\text{s}\times m'}\), and the corresponding transmitted symbols are represented by \(\mathbf{X}_{k,i}\in\mathbb{C}^{N_\text{s}\times m'}\).\par

However, dividing data into multiple blocks reduces the number of symbols available for semi-blind CE. In high-order modulation like 256-QAM, the insufficient number of samples may prevent the optimization problem from achieving the optimal solution. 
A data augmentation method is proposed based on the symmetry of the \textit{M}-QAM constellation to increase the number of samples available for semi-blind CE. 
Denote \(\mathcal{X}=\{ {x_1},{x_2},\ldots,{x_M}\}\) as the \textit{M}-QAM constellation, for \(x_i\in\mathcal{X}_M\), we have \(-x_i\in\mathcal{X}_M\), \(jx_i\in\mathcal{X}_M\) and \(-jx_i\in\mathcal{X}_M\).
Consider two transmitted symbol vectors \(\mathbf{x}_{k1}\in\mathbb{C}^{N_\text{s}}\) and \(\mathbf{x}_{k2}\in\mathbb{C}^{N_\text{s}}\), such that \(\mathbf{x}_{k1}=-\mathbf{x}_{k2}\), and \(\mathbf{x}_{k1}\) is drawn from \(\mathcal{X}_M\), they satisfy
\begin{equation}
    \label{e21}
    \setlength\abovedisplayskip{3pt}
    \setlength\belowdisplayskip{3pt}
    \begin{split}
        p(\mathbf{y}_{k1}|\mathbf{x}_{k2})&=\mathbb{P}(\mathbf{y}_{k1}=-\mathbf{H}_k^{\text{eq}}\mathbf{x}_{k1}+\mathbf{\widetilde n}_k)\\
        &=\mathbb{P}(-\mathbf{y}_{k1}=\mathbf{H}_k^{\text{eq}}\mathbf{x}_{k1}-\mathbf{\widetilde n}_k)\\
        &=\mathbb{P}(-\mathbf{y}_{k1}=\mathbf{H}_k^{\text{eq}}\mathbf{x}_{k1}+\mathbf{\widetilde n}_k)\\
        &=p(-\mathbf{y}_{k1}|\mathbf{x}_{k1}),
    \end{split}
\end{equation}
where \(\mathbf{y}_{k1}\in\mathbb{C}^{N_\text{s}}\) is the corresponding received symbol vector of \(\mathbf{x}_{k1}\). Similarly, we can prove that
\begin{equation}
    \label{e22}
    \setlength\abovedisplayskip{3pt}
    \setlength\belowdisplayskip{3pt}
    \begin{split}
        &p(\mathbf{y}_{k1}|\mathbf{x}_{k3})=p(j\mathbf{y}_{k1}|\mathbf{x}_{k1}), \text{for}\;\mathbf{x}_{k3}=-j\mathbf{x}_{k1},\\
        &p(\mathbf{y}_{k1}|\mathbf{x}_{k4})=p(-j\mathbf{y}_{k1}|\mathbf{x}_{k1}), \text{for}\;\mathbf{x}_{k4}=j\mathbf{x}_{k1}.
    \end{split}
\end{equation}

Therefore, \(\mathbf{y}_{k2}=-\mathbf{y}_{k1}\), \(\mathbf{y}_{k3}=j\mathbf{y}_{k1}\) and \(\mathbf{y}_{k4}=-j\mathbf{y}_{k1}\). Based on the symmetry of the \textit{M}-QAM constellation, \(\mathbf{x}_{k2}\), \(\mathbf{x}_{k3}\), \(\mathbf{x}_{k4}\) are also drawn from \(\mathcal{X}_M\). Based on this property, we adopt the following data augmentation method to increase the number of samples available for semi-blind CE. 
For received symbols \(\mathbf{Y}_{k, i}\), let \(\mathbf{Y}_{k, i}^{(1)}=-\mathbf{Y}_{k, i}\), \(\mathbf{Y}_{k, i}^{(2)}=j\mathbf{Y}_{k, i}\) and \(\mathbf{Y}_{k, i}^{(3)}=-j\mathbf{Y}_{k, i}\), the augmented received symbol matrix is given by \(\mathbf{\widetilde{Y}}_{k, i}\triangleq [\mathbf{Y}_{k, i},\mathbf{Y}_{k, i}^{(1)},\mathbf{Y}_{k, i}^{(2)},\mathbf{Y}_{k, i}^{(3)}]\), thereby increasing the number of received symbols available for semi-blind CE by up to four times the original amount.\par

It is important to note that the data augmentation method described above does not introduce additional information. 
The \textit{M} in (\ref{e19}) refers to the boundary of the constellation, ensuring that the estimated symbols are close to the expected constellation points.
Data augmentation aims to increase the number of samples positioned at the boundary of the constellation without altering the boundary itself. 
This adjustment does not affect the solution of the optimization problem while mitigating the issue of insufficient samples at the boundary of the constellation in high-order modulation.
Such a deficiency could render the constraint in (\ref{e19}) ineffective.\par



\begin{algorithm}[htbp]
    \caption{Modified Semi-Blind JCESD for Frequency-Selective Channels}
    \label{alg2}
    \renewcommand{\algorithmicrequire}{\textbf{Input:}}
    \renewcommand{\algorithmicensure}{\textbf{Output:}}
    \begin{algorithmic}[1]
        \REQUIRE Received signals \({\mathbf{Y}_k}[j]\).
        \ENSURE An estimate of the channel matrix \({\hat{\mathbf{H}}_{k,N_{\text{iter}}}^{\text{eq}}[j]}\), and an estimate of the transmitted symbols \({\mathbf{\hat X}_{k,N_{\text{iter}}}[j]}\) after \(N_\text{iter}\) iterations.
        \FOR{\(i=1:N_f\)}
            \STATE \(\mathbf{Y}_{k,i} \triangleq [ {{{\mathbf{Y}}_k}[i \times \Delta f], \ldots ,{{\mathbf{Y}}_k}[(i + 1) \times \Delta f]}]\)
            \STATE Normalize \(\mathbf{Y}_k\) via (\ref{e20})
            \STATE \(\mathbf{\widetilde{Y}}_{k,i} \triangleq [\mathbf{Y}_{k,i}, -\mathbf{Y}_{k,i}, j\mathbf{Y}_{k,i}, -j\mathbf{Y}_{k,i}]\)
            \IF{\(\kappa({\mathbf{\widetilde{Y}}_{k,i}}) > \kappa_\text{max}\)}
                \RETURN FAIL
            \ENDIF
            \STATE \(\mathbf{U}_0={\left( {{\mathbf{\hat H}}_k^{{\text{eq}}}} \right)^{ - 1}}\) via (\ref{e42})
            \WHILE{\({\left\| {{{\mathbf{U}}_0}{\mathbf{\widetilde{Y}}}_{k,i}} \right\|_\infty } > 1\)}
                \STATE \(\mathbf{U}_0=\mathbf{U}_0/2\)
            \ENDWHILE
            \STATE Perform SLSQP over (\ref{e18}), (\ref{e30}) and (\ref{e36}) starting at \(\mathbf{U}_0\) to find an optimal value \(\mathbf{U}_{k,i}\)
            \STATE \(\mathbf{\hat{H}}_k^{\text{eq}}[i\times\Delta f:(i+1)\times\Delta f]=\mathbf{U}_{k,i}^{-1}\)
            \STATE \(\mathbf{\hat{X}}_k[i\times\Delta f:(i+1)\times\Delta f]=\lfloor \mathbf{U}_{k,i}\mathbf{Y}_k[i\times\Delta f:(i+1)\times\Delta f]\rceil\)
            \STATE Eliminate the ambiguity
        \ENDFOR
        \FOR{\(t=0:N_{\text{iter}}\)}
            \IF{\(t=0\)}
                \STATE \(\mathbf{\hat{X}}_{k,t}[j]=\mathbf{\hat{X}}_k[j]\)
            \ENDIF
            \STATE Symbols in \(\mathbf{\hat{X}}_{k,t}[j]\) with \(llr<\lambda\) are removed to obtain \(\mathbf{\hat{X}}_{k,t}'[j]\)
            \STATE \(\mathbf{\hat{H}}_{k,t+1}[j]=\left(\mathbf{\hat X}'^{\mathsf{H}}_{k,t}[j]\mathbf{\hat{X}}'_{k,t}[j]\right)^{-1}\mathbf{\hat X}'_{k,t}[j]\mathbf{Y}_k[j]\)
            \STATE Obtain \(\mathbf{\hat{X}}_{k,t+1}[j]\) via (\ref{e39}) and (\ref{e40})
        \ENDFOR
        \RETURN \({\hat{\mathbf{H}}_{k,N_{\text{iter}}}^{\text{eq}}[j]}\), \({\mathbf{\hat X}_{k,N_{\text{iter}}}[j]}\).
    \end{algorithmic}
\end{algorithm}

\subsubsection{Iterative JCESD}
Considering the accuracy of semi-blind JCESD is relatively low, we adopt the following iterative JCESD method. By progressively reducing the symbol error rate (SER) in each iteration, the accuracy of CE is improved. \par
For the \(t^{\text{th}}\) iteration, denote \(\mathbf{\hat X}_{k,t}[j]\in\mathbb{C}^{N_\text{s}\times T}\) as the input. For the first iteration (\(t=0\)), \(\mathbf{\hat X}_{k,0}[j]\) is obtained from the output of the semi-blind CE. Based on LS estimation, the channel matrix at the next iteration is given by 
\begin{equation}
    \label{e37}
    \setlength\abovedisplayskip{3pt}
    \setlength\belowdisplayskip{3pt}
    \mathbf{\hat H}_{k,t+1}[j]=\left(\mathbf{\hat X}^{\mathsf{H}}_{k,t}[j]\mathbf{\hat X}_{k,t}[j]\right)^{-1}\mathbf{\hat X}_{k,t}[j]\mathbf{Y}_k[j],
\end{equation}
in the process above, \(\mathbf{\hat X}_{k,t}[j]\) serves a role analogous to pilot symbols.
To mitigate the effect of SD errors, its elements are projected onto the nearest constellation points before use.
Unlike the conventional pilot-based CE method, this approach leverages all available symbols on each subcarrier for LS estimation. \par
However, under conditions of severe noise and IUI, this hard decision mechanism may introduce erroneous samples.
Incorporating these unreliable samples into LS estimation can lead to significant errors in \(\mathbf{\hat H}_{k,t+1}[j]\). 
To address this, we exclude samples with high error probabilities from the estimation process.
Specifically, for a given sample, let the Euclidean distances to the four closest points on the constellation be ordered from smallest to largest as \(d_1\), \(d_2\), \(d_3\) and \(d_4\), define its log-likelihood ratio (LLR) as
\begin{equation}
    \label{e38}
    \setlength\abovedisplayskip{3pt}
    \setlength\belowdisplayskip{3pt}
    {\text{llr}} = \ln \frac{{\exp \left\{ { - \frac{{{d_1}}}{{2\sigma _{k,x}^2}}} \right\}}}{{\sum\nolimits_{i = 2}^4 {\exp \left\{ { - \frac{{{d_i}}}{{2\sigma _{k,x}^2}}} \right\}} }},
\end{equation}
where \(\sigma_{k,x}^2\) is derived from the SINR of \(\mathbf{\hat X}_{k,t}[j]\). For practical implementation, we approximate this value as \(\sigma_{k,x}^2=1/\text{SNR}_k\). In high SNR scenarios, since \(\sigma_{k,x}^2\) is relatively small, to avoid numerical overflow, we impose a lower bound \(\sigma_{k,x}^2=\max\{{10^{-3}}, 1/\text{SNR}_k\}\).
If the computed LLR of the sample falls below a predefined decision threshold \(\lambda\), the sample is deemed unreliable and discarded before LS estimation. This selective filtering effectively retains only the samples with small minimum Euclidean distances, reducing the SER of the retained samples and consequently enhancing the accuracy of the CE. \par

After refining the channel matrix, linear minimum mean square error (LMMSE) SD is performed to obtain a more accurate \(\mathbf{\hat X}_{k,t+1}[j]\). The SD process is
\begin{equation}
    \label{e39}
    \setlength\abovedisplayskip{3pt}
    \setlength\belowdisplayskip{3pt}
    \mathbf{G}_{k,t+1}[j] = \mathbf{\hat H}_{k,t+1}^{\mathsf{H}}[j]{(\mathbf{\hat H}_{k,t+1}[j]\mathbf{\hat H}_{k,t+1}^{\mathsf{H}}[j] + {\sigma_k^2}{\mathbf{I}})^{-1}},
\end{equation}
\begin{equation}
    \label{e40}
    \setlength\abovedisplayskip{3pt}
    \setlength\belowdisplayskip{3pt}
    \mathbf{\hat X}_{k,t+1}[j]=\text{diag}(\mathbf{G}_{k,t+1}[j]\mathbf{\hat H}_{k,t+1})^{-1}\mathbf{G}_{k,t+1}[j]\mathbf{Y}_k[j].
\end{equation}

Through \(N_{\text{iter}}\) iterations, the process will converge to refined estimates of both \(\mathbf{\hat X}_{k,t}[j]\) and \(\mathbf{\hat H}_{k,t}[j]\). The complete iterative JCESD procedure is summarized in Algorithm \ref{alg2}.

\subsection{Computational Complexity}
Calculation of the optimization program in (\ref{e18}) consists of two parts. The first part is the determinant gradient, which is
\begin{equation}
    \setlength\abovedisplayskip{3pt}
    \setlength\belowdisplayskip{3pt}
    \nabla \left(\log \left| \det(\mathbf{U}) \right|\right) = \left(\mathbf{U}^{-1} \right)^{\mathsf{T}},
    \label{e421}
\end{equation}
with a computational complexity of order \(\mathcal{O}(N_\text{s}^3)\). Another part is the constraint check in (\ref{e30}), which consists of matrix multiplications with complexity \(\mathcal{O}(N_\text{s}^2m')\). So the computational complexity per TTI is
\begin{equation}
    \label{e43}
    \setlength\abovedisplayskip{3pt}
    \setlength\belowdisplayskip{3pt}
    \mathcal{C}_{\text{opt}}=N_\text{f}\times T_{\text{iter}}\times\mathcal{O}(N_\text{s}^3+N_\text{s}^2m'),
\end{equation}
where \(T_{\text{iter}}\) denotes the average number of iterations required to solve the optimization problem each time. \par
The computational process in the iterative JCESD can also be divided into two parts. The first part is LS estimation in (\ref{e37}), which involves three matrix multiplications and one matrix inversion, with a computational complexity of \(\mathcal{O}(3N_\text{s}^3+N_\text{s}^2m')\). The second part is LMMSE equalization in (\ref{e39}) and (\ref{e40}), which involves five matrix multiplications and two matrix inversions, with a computational complexity of \(\mathcal{O}(6N_\text{s}^3+N_\text{s}^2m')\). So the computational complexity per TTI of the JCESD process is
\begin{equation}
    \label{e44}
    \setlength\abovedisplayskip{3pt}
    \setlength\belowdisplayskip{3pt}
    \mathcal{C}_{\text{iter}} = J\times N_{\text{iter}}\times\mathcal{O}(9N_\text{s}^3+2N_\text{s}^2m').
\end{equation}
\par
Compared with the conventional pilot-based methods, semi-blind JCESD provides better throughput (see Section \ref{sec4}) but at the cost of higher computational complexity.

\subsection{Pilot Overhead and Theoretical Gain}

\begin{figure}[h]
    \centering
    \setlength{\abovecaptionskip}{0.cm}
    \subfigure[Orthogonal Pilots]{
        \centering
        \includegraphics[width=0.22\textwidth]{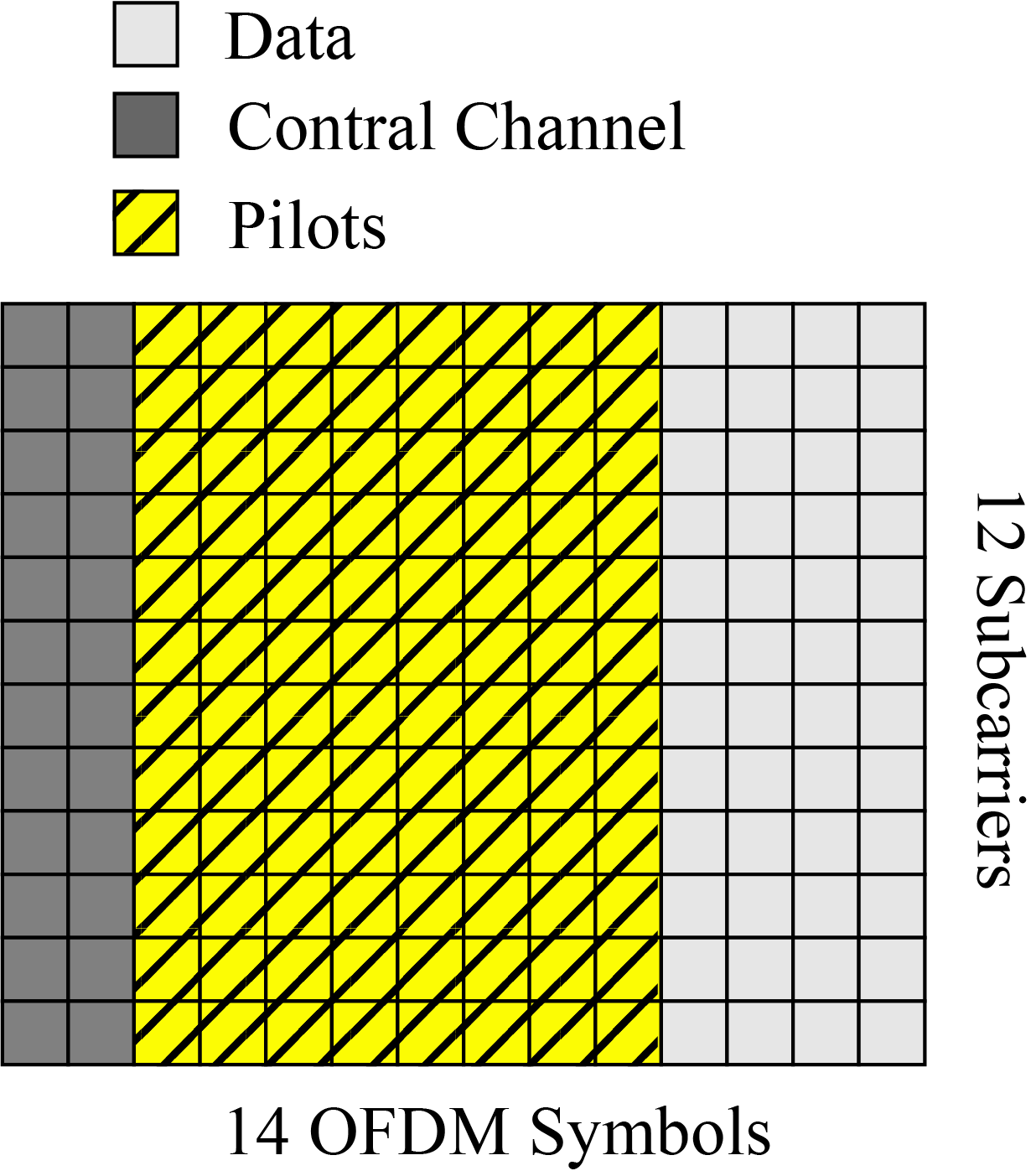}
        \label{fig6a}
    }
    \subfigure[Non-Orthogonal Pilots]{
        \centering
        \includegraphics[width=0.22\textwidth]{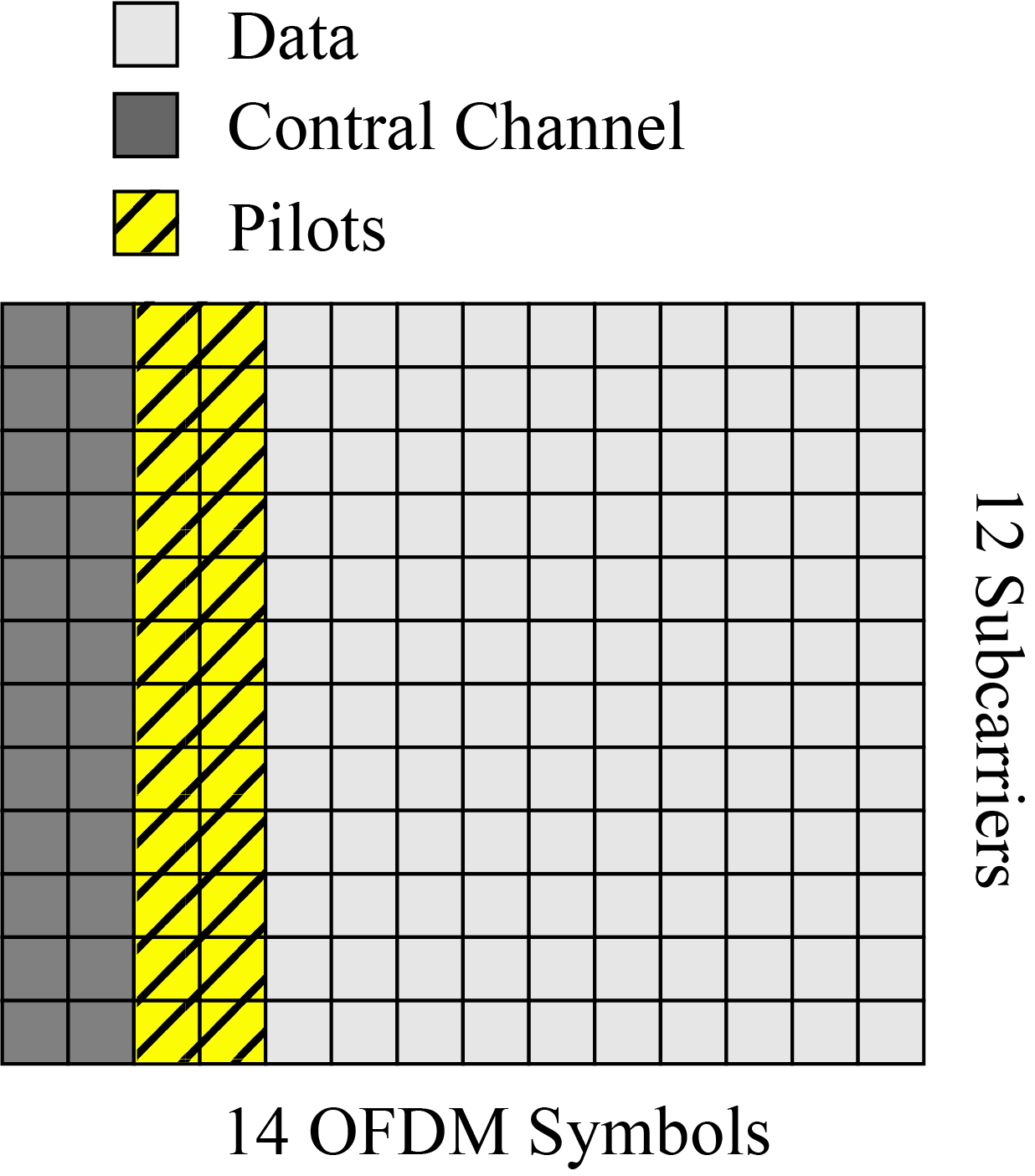}
        \label{fig6b}
    }
    \caption{The pilot pattern specified by the 3GPP standard with 96 streams}
    \label{fig6}
    \vspace{-0.3cm}
\end{figure}

Fig. \ref{fig6} illustrates the pilot overhead incurred by the pilot-based methods widely adopted in current 3GPP standards when transmitting 96 data streams \cite{ref35}. Under the constraint of not increasing the number of OFDM symbols allocated for pilot transmission, the system can support at most 24 spatial streams with fully orthogonal pilot sequences. As the number of users increases, the overhead becomes increasingly significant. Specifically, in the case of transmitting 96 data streams with orthogonal pilots, a total of 96 resource elements (REs) are occupied by pilot symbols, accounting for approximately 67\% of the available REs in the corresponding resource block. In contrast, the proposed JCESD method significantly reduces the pilot overhead. For example, with \(N_\text{s}=\)2 , only 2 REs are required for pilot transmission regardless of the total number of users. Under the same modulation order and code rate, this results in a theoretical maximum throughput gain of up to 196\% compared to the conventional fully orthogonal pilot scheme. \par

Furthermore, in a practical scenario where no additional OFDM symbols are introduced for pilot transmission, the pilot overhead is constrained to 17\% of the total REs. The proposed JCESD method achieves a theoretical maximum throughput gain of around 20\% when \(N_\text{s}=\)2. This substantial improvement demonstrates the key advantage of semi-blind JCESD: by minimizing pilot overhead without sacrificing accurancy, it enables more efficient use of time-frequency resources, thereby enhancing overall system throughput. \par

\section{Simulation Results}
\label{sec4}
This section evaluates the performance of the proposed algorithms and compares the simulation results with pilot-based CE methods specified in the Third Generation Partnership Project (3GPP) standard. The results are derived from simulations conducted over 20 consecutive TTIs to assess the average normalized mean squared error (NMSE), bit error ratio (BER), and throughput as a function of SNR. The NMSE is defined as
\begin{equation}
    \label{e41}
    \setlength\abovedisplayskip{3pt}
    \setlength\belowdisplayskip{3pt}
    {\text{NMSE}}=\frac{1}{{J\times K}}\sum\limits_{k=1}^K{\sum\limits_{j=1}^J {\frac{{\left\|{{{{\mathbf{\hat H}}}_k^{\text{eq}}}[j]-{\mathbf{H}}_k^{{\text{eq}}}[j]}\right\|_\mathsf{F}^2}}{{\left\|{{\mathbf{H}}_k^{{\text{eq}}}[j]}\right\|_\mathsf{F}^2}}}}.
\end{equation}

Throughput is defined as the total number of correctly received bits across all users and TTIs. A block error model is adopted, where if an error is detected within a data block, all bits in that block are considered erroneous. The simulations are conducted using the platform described in \cite{ref11}. \par

\subsection{Simulation Setup}

The parameters used in the simulation are summarized in Tab. \ref{t1}. The antenna elements of the BS's antenna array are arranged in a 48-row by 16-column UPA and operate in a dual-polarized omnidirectional configuration. All users are uniformly distributed around the BS. \par

\begin{table}[t]
    \caption{System Parameter Configurations}
    \label{t1}
    \centering
    \begin{tabular}{cc}
        \toprule
        \textbf{Parameter} & \textbf{Specification} \\
        \midrule
         Channel model & CDL-C \\
         Delay spread & 100 ns \\
         Maximum speed of users & 3.0 km/h \\
         Carrier frequency & 6.7 GHz \\ 
         Subcarrier space & 30 kHz \\
         SNR & 0 $\sim$ 27 dB \\
         Number of users&\(K=\) 24, 48 \\
         Number of data streams per user& \(N_\text{s}=\) 2\\
         Number of receive RF chains & \(N_\text{r}^\text{RF}=\) 16 \\
         Number of transmit RF chains & \(N_\text{t}^\text{RF}=\) 256 \\
         Number of transmit antennas & \(N_\text{t}=\) 1536 \\
         Number of OFDM subcarriers & \(J=\) 48 \\ 
         Number of OFDM symbols & \(Y=\) 14 \\
         Number of TTIs & 20 \\
         Number of data blocks & \(N_\text{f}=\) 8 \\
         Number of iterations & \(N_\text{iter}=\) 5 \\
         Decision threshold  & \(\lambda=\) 15 \\
         \bottomrule
    \end{tabular}
\end{table}

\vspace{-0.3cm}
\begin{table}[h]
    \caption{MCS Index Table}
    \label{t2}
    \centering
    \begin{tabular}{ccc}
        \toprule
        MCS Index&Modulation Order&LDPC Code Rate \\
        \midrule
        5&4&0.396 \\
        10&4&0.643 \\
        11&6&0.455 \\
        19&6&0.853 \\
        20&8&0.667 \\
        27&8&0.926 \\
        \bottomrule
    \end{tabular}
    \vspace{-0.3cm}
\end{table}

To maximize throughput, a link adaptation technique is adopted to select the modulation and coding scheme (MCS) with the highest index, as shown in Tab. \ref{t2}, while ensuring that the block error rate (BLER) remains below 0.1 (a cyclic redundancy check is employed to determine the presence of errors in each data block). It is worth noting that the 3GPP standard defines a total of 28 MCS levels \cite{ref35}, while Tab. \ref{t2} lists only a representative subset. 
Three methods are comparatively evaluated in this scenario. 
\begin{itemize}
    \item Pilot-based methods, including two variants: \textit{Orthogonal Pilots}, which use additional OFDM symbols to ensure pilot orthogonality, and \textit{Non-Orthogonal Pilots}, where pilot reuse is allowed among users.
    Pilot sequences follow the design in \cite{ref35} and the frequency-domain Wiener filter from \cite{ref12} is used to interpolate CE results across the band.
    \item Semi-blind method, where the EM-based algorithm from \cite{ref39} is used, and LMMSE detection is implemented as in (\ref{e39}) and (\ref{e40}).
    \item Blind method, represented by the approach in \cite{ref1}.
\end{itemize}

\subsection{Throughput}
This subsection presents simulation results used to analyze the throughput improvement achieved by the proposed algorithm. The proposed method is compared against three baselines: EM-based semi-blind CE \cite{ref39}, and pilot-based CE with orthogonal and non-orthogonal pilot patterns defined in \cite{ref35}. \par

\begin{figure*}[htbp]
    \subfigure[\textit{K} = 24]{
        \label{fig7a}
        \centering
        \includegraphics[width=0.48\textwidth]{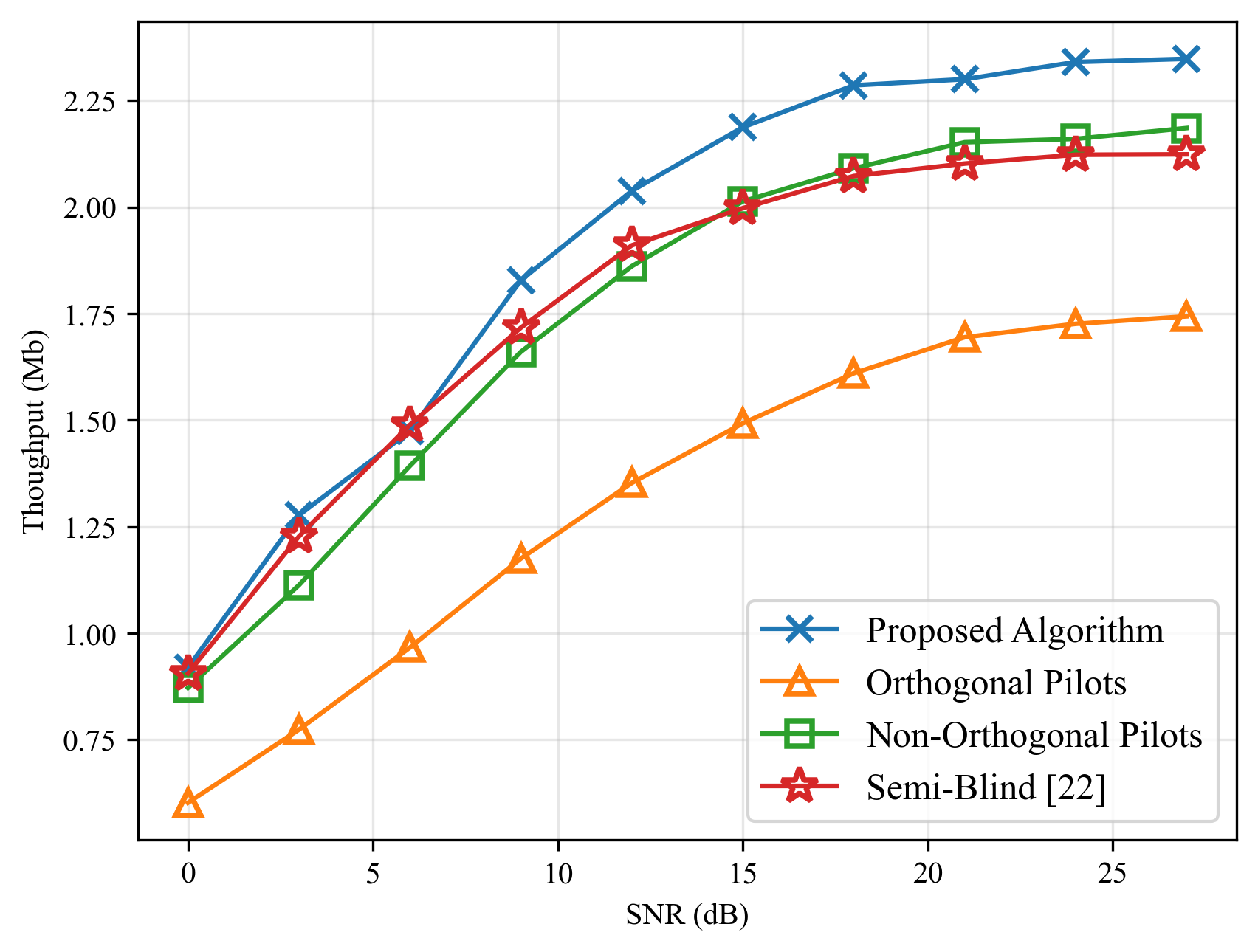}
    }
    \subfigure[\textit{K} = 48]{
        \centering
        \label{fig7b}
        \includegraphics[width=0.48\textwidth]{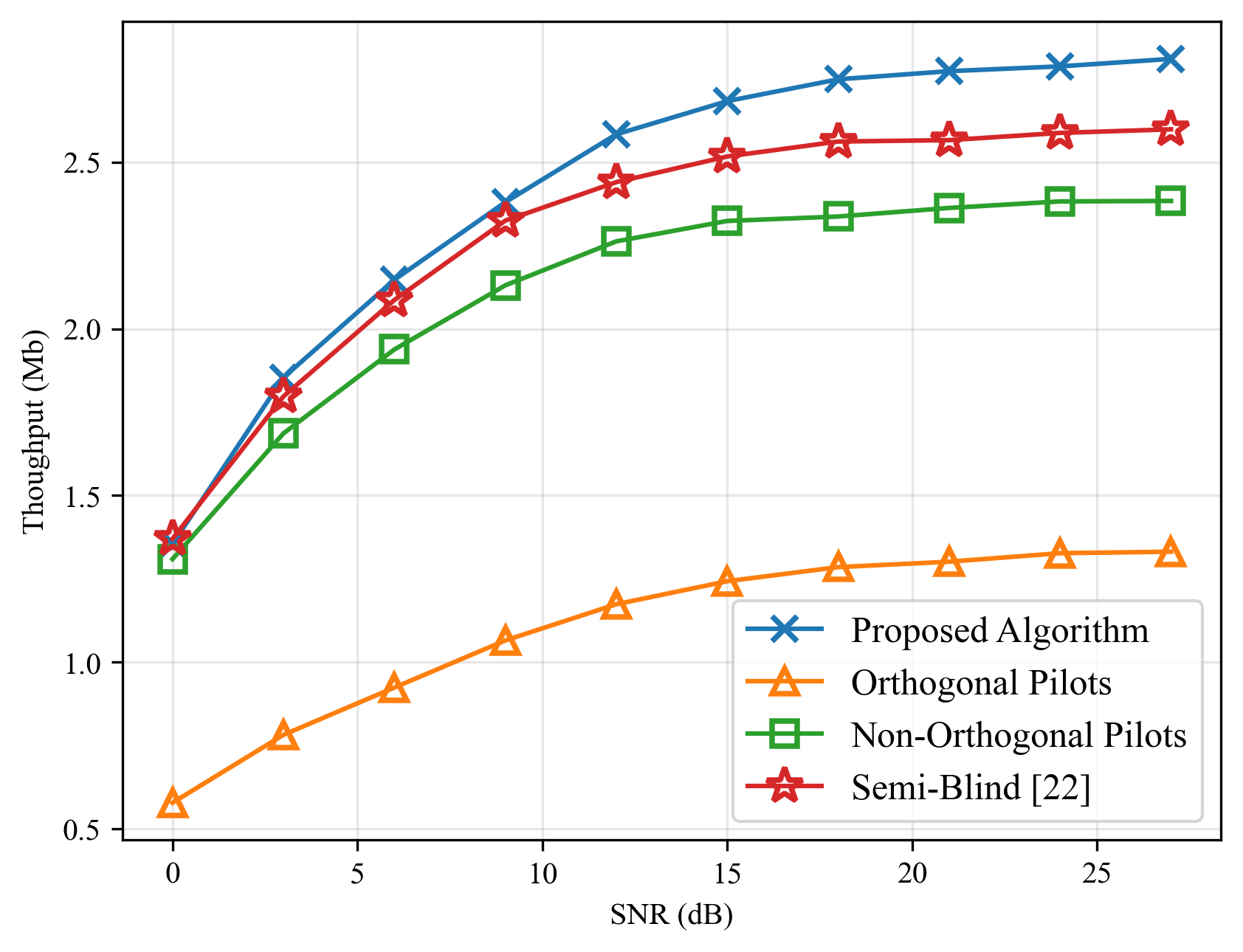}
    }
    \centering
    \caption{The throughput of the proposed algorithm: (a) 24 users transmitting 48 streams, (b) 48 users transmitting 96 streams.} 
    \label{fig7}
    \vspace{-0.5cm}
\end{figure*}

Fig. \ref{fig7} illustrates the throughput performance of the proposed algorithm compared with pilot-based methods under two scenarios: 24 and 48 users. 
In both cases, the proposed algorithm consistently achieves higher throughput across all SNR levels, with diminishing growth at high SNRs due to IUI becoming dominant over AWGN.
Notably, throughput gain plateaus beyond 18 dB SNR, reflecting the limited impact of increased SNR when interference power remains unchanged.
In the 24-user scenario, the pilot-based methods incur a significant overhead, which the proposed semi-blind JCESD method avoids. 
However, since pilot-based methods can exploit pilot symbols for IUI suppression during SD, the proposed method may experience performance degradation, especially with higher modulation orders, where minor SINR increases may not correspond to MCS improvements due to decreased accuracy.
In the 48-user scenario, the IUI becomes more severe. Orthogonal pilot-based methods require six additional OFDM symbols, further amplifying overhead. Despite doubling the number of users, the throughput does not scale proportionally, particularly at high SNRs. 
The proposed algorithm shows improved resilience to IUI in this case, achieving up to 20\% throughput gain over non-orthogonal pilot-based methods, compared to less than 15\% in the 24-user case, indicating its robustness in interference-limited environments.

\vspace{-0.3cm}
\begin{table}[h]
    \centering
    \caption{The Throughput Gain (\%) of Proposed Algorithm}
    \begin{tabular}{ccccc}
        \toprule
        \multirow{2}{*}{SNR (dB)} & \multicolumn{2}{c}{Orthogonal Pilot} & \multicolumn{2}{c}{Non-Orthogonal Pilot} \\
        \cline{2-5}
        & \(K=\)24 & \(K=\)48 & \(K=\)24 & \(K=\)48 \\
        \midrule
        0 & 27.3 & 134.1 & 5.5 & 3.4 \\
        3 & 37.5 & 137.0 & 14.7 & 9.9 \\
        6 & 27.1 & 132.4 & 5.8 & 10.8 \\
        9 & 29.6 & 123.4 & 10.1 & 11.7 \\
        12 & 25.6 & 120.1 & 9.5 & 14.2 \\
        15 & 22.1 & 115.9 & 8.6 & 15.5 \\
        18 & 18.3 & 113.8 & 9.3 & 17.6 \\
        21 & 12.1 & 113.1 & 6.9 & 17.4 \\
        24 & 13.0 & 110.3 & 8.3 & 17.0 \\
        27 & 12.2 & 111.0 & 7.4 & 17.9 \\
        \bottomrule
    \end{tabular}
    \label{tab3}
\end{table}

\subsection{CE Accuracy}

\begin{figure*}[ht]
    \subfigure[\textit{K} = 24]{
        \centering
        \label{fig10a}
        \includegraphics[width=0.48\textwidth]{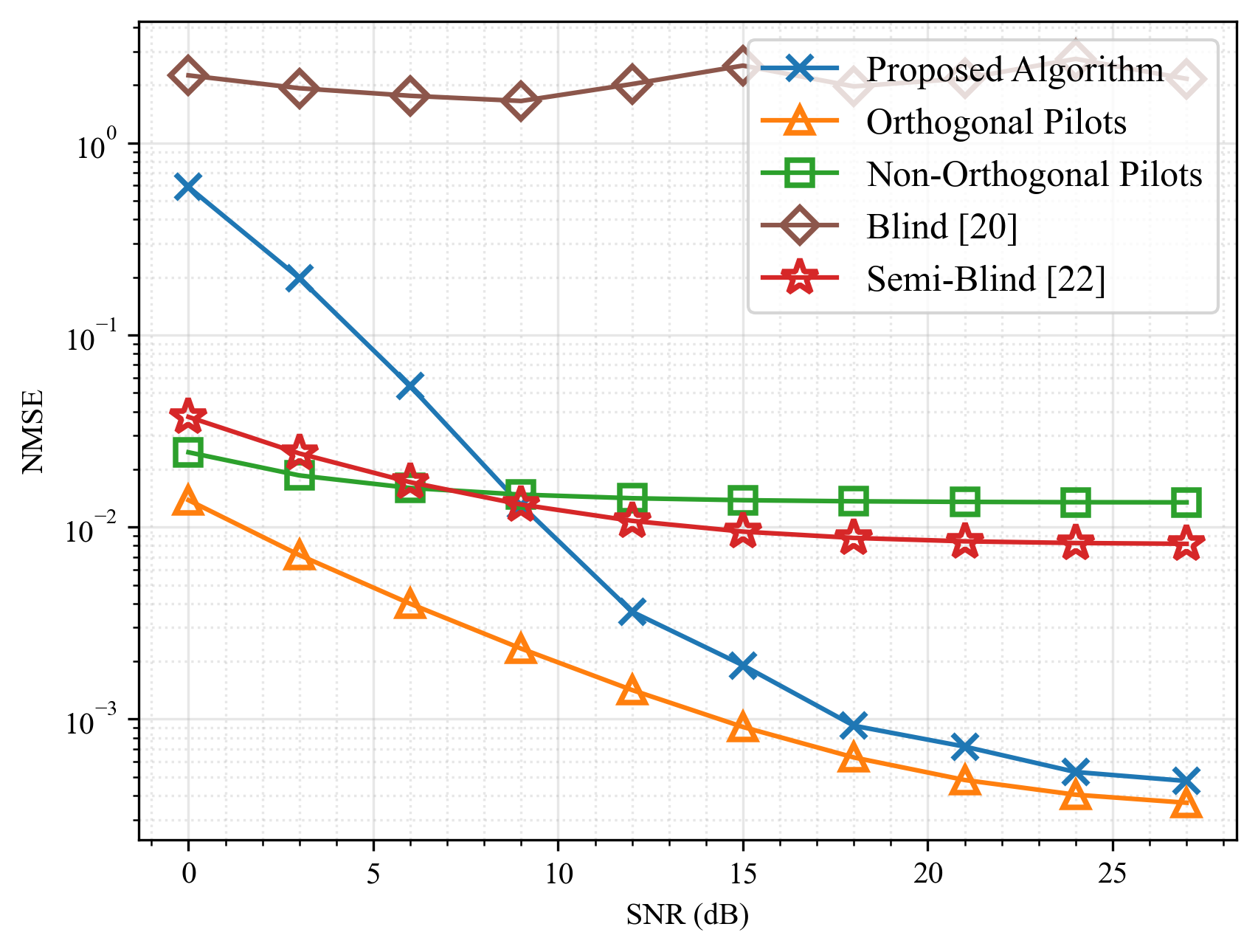}
    }
    \subfigure[\textit{K} = 48]{
        \centering
        \label{fig10b}
        \includegraphics[width=0.48\textwidth]{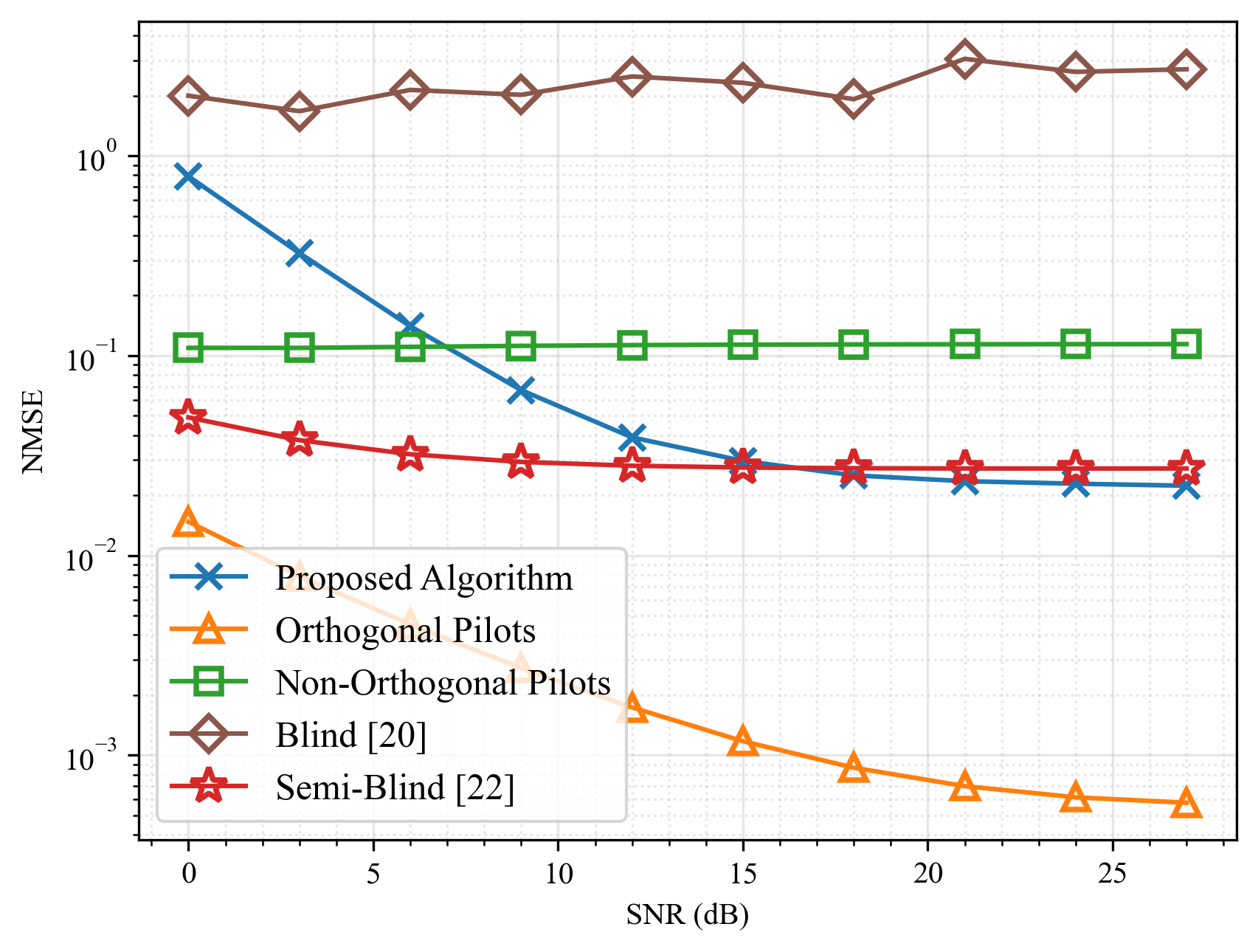}
    }
    \centering
    \setlength{\abovecaptionskip}{0.cm}
    \caption{The NMSE of the proposed algorithm: (a) 24 users transmitting 48 streams, (b) 48 users transmitting 96 streams.} 
    \label{fig10}
    \vspace{-0.4cm}
\end{figure*}

\begin{figure*}[ht]
    \subfigure[\textit{K} = 24]{
        \label{fig9a}
        \centering
        \includegraphics[width=0.48\textwidth]{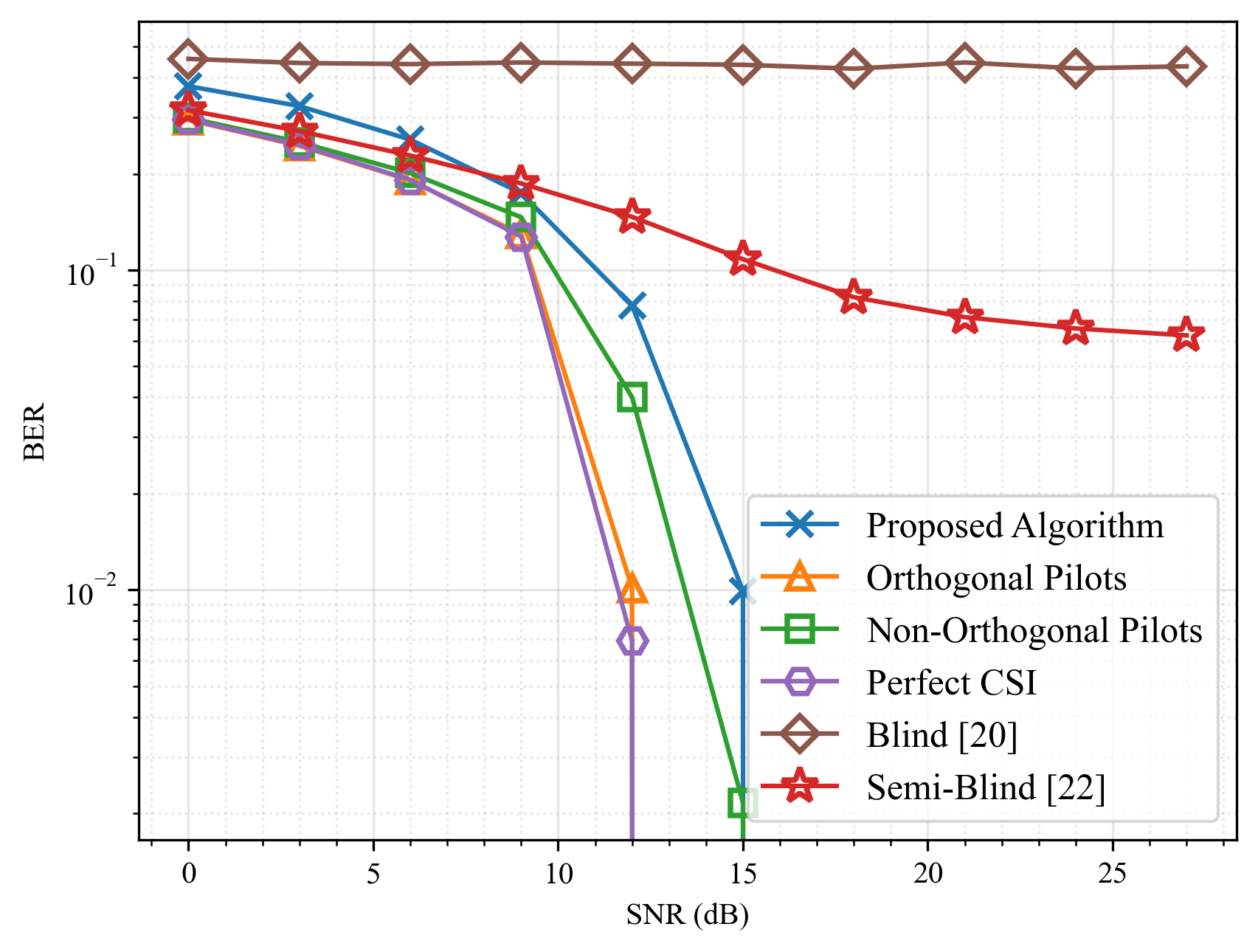}
    }
    \subfigure[\textit{K} = 48]{
        \centering
        \label{fig9b}
        \includegraphics[width=0.48\textwidth]{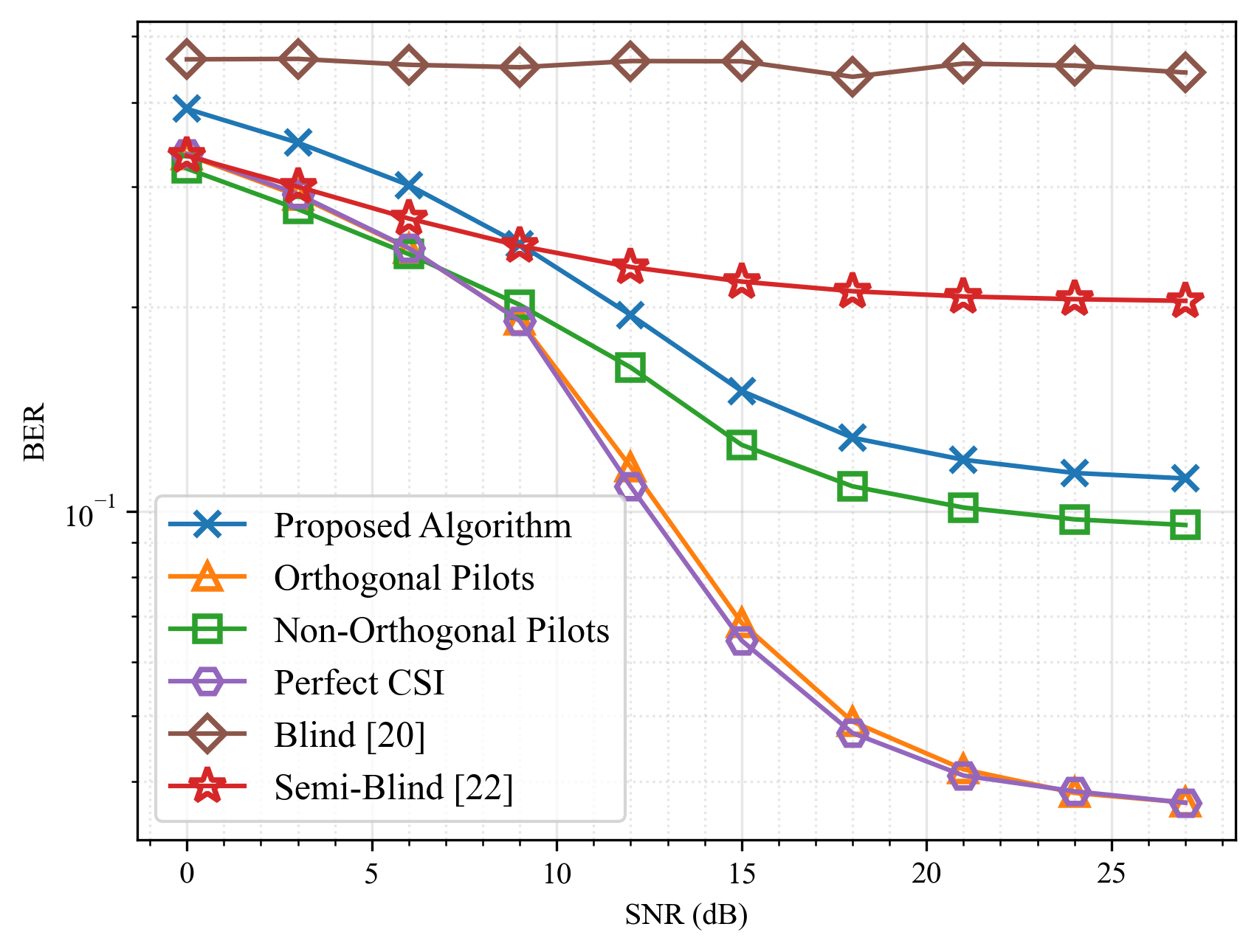}
    }
    \centering
    \setlength{\abovecaptionskip}{0.cm}
    \caption{The BER of the proposed algorithm: (a) 24 users transmitting 48 streams, (b) 48 users transmitting 96 streams.} 
    \label{fig9}
    \vspace{-0.4cm}
\end{figure*}

This subsection evaluates the estimation accuracy of the proposed algorithm in terms of NMSE and average BER across all users. 
In addition to the orthogonal and non-orthogonal pilot-based methods discussed previously, comparisons are also made with the algorithms presented in \cite{ref1} and \cite{ref39}. 
For the BER simulation, all users are configured to use the MCS with index of 20 from Tab. \ref{t2}, corresponding to 256-QAM and an LDPC code rate of 0.667. 
Additionally, for the BER comparison, a perfect CSI baseline is considered, assuming that users have full knowledge of the channel. \par

Fig. \ref{fig10} illustrates the NMSE performance of the evaluated algorithms as a function of SNR.
Since the method in \cite{ref1} does not support high-order modulation, it fails under the 256-QAM scenario as evidenced by significantly high values of both BER and NMSE.
When the number of users is small and IUI is limited, the proposed algorithm achieves CE accuracy comparable to pilot-based methods at high SNR, despite utilizing only a small number of pilot symbols.
Moreover, despite operating under a significantly higher modulation order than the method in \cite{ref1}, the proposed approach achieves even better estimation accuracy.
Owing to severe IUI among pilot symbols, the NMSE of the non-orthogonal pilot-based method remains largely unaffected by increasing SNR, resulting in consistently low estimation accuracy.
In scenarios with a higher user density, increased IUI causes degradation in estimation accuracy across all methods except the orthogonal pilot-based scheme.
The use of high-order modulation in the proposed algorithm reduces its resilience to noise and IUI, resulting in a more pronounced degradation in estimation accuracy.
Nevertheless, the proposed algorithm still outperforms the non-orthogonal pilot-based method in terms of estimation accuracy. \par

Fig. \ref{fig9} presents the average BER across all users for the evaluated algorithms.
The orthogonal pilot-based method achieves high CE accuracy, resulting in a BER that closely approaches the performance under perfect CSI.
However, the excessive pilot overhead prevents the low BER of this method from translating into higher throughput when compared to the proposed algorithm.
Although the proposed algorithm achieves higher CE accuracy than the non-orthogonal pilot-based method, the latter benefits from pilot symbols to partially suppress IUI, thereby achieving a lower BER in some scenarios.
While the BER of the proposed algorithm is not as favorable as that of pilot-based schemes, the substantial reduction in pilot overhead leads to significantly improved throughput.
In low-IUI scenarios, as the SNR increases, the BER of the proposed algorithm converges to that of the pilot-based methods. \par

\begin{figure}[h]
    \includegraphics[width=0.48\textwidth]{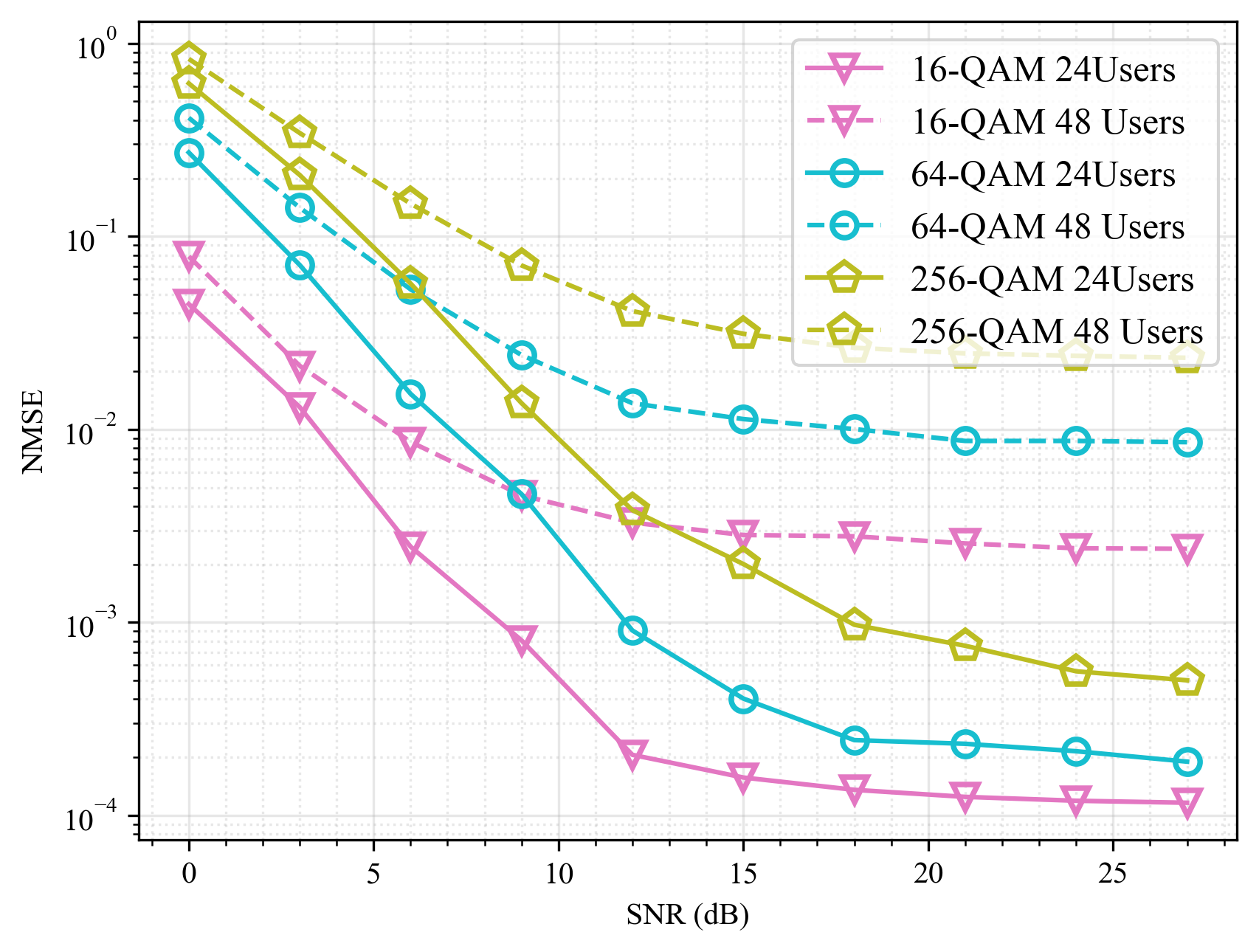}
    \centering
    \setlength{\abovecaptionskip}{0.cm}
    \caption{The NMSE of the proposed algorithm under different modulation orders.} 
    \label{fig11}
\end{figure}

As the proposed algorithm performs CESD by fitting the received symbols to the \textit{M}-QAM constellation, its CE accuracy is inherently dependent on the modulation order, as depicted in Fig. \ref{fig11}. 
As SNR increases, the NMSE of the algorithm converges to a minimum value that varies with the modulation order. 
Higher modulation orders lead to a corresponding increase in the minimum achievable NMSE. 
Under 16-QAM and 64-QAM, the proposed algorithm achieves a lower minimum NMSE compared to the orthogonal pilot-based method, as shown in Fig. \ref{fig10}. 
As indicated by the dashed segments in Fig. \ref{fig11}, increasing the modulation order by one level at a fixed SINR results in approximately a 5 dB degradation in CE accuracy. \par

\begin{figure}[h]
    \includegraphics[width=0.48\textwidth]{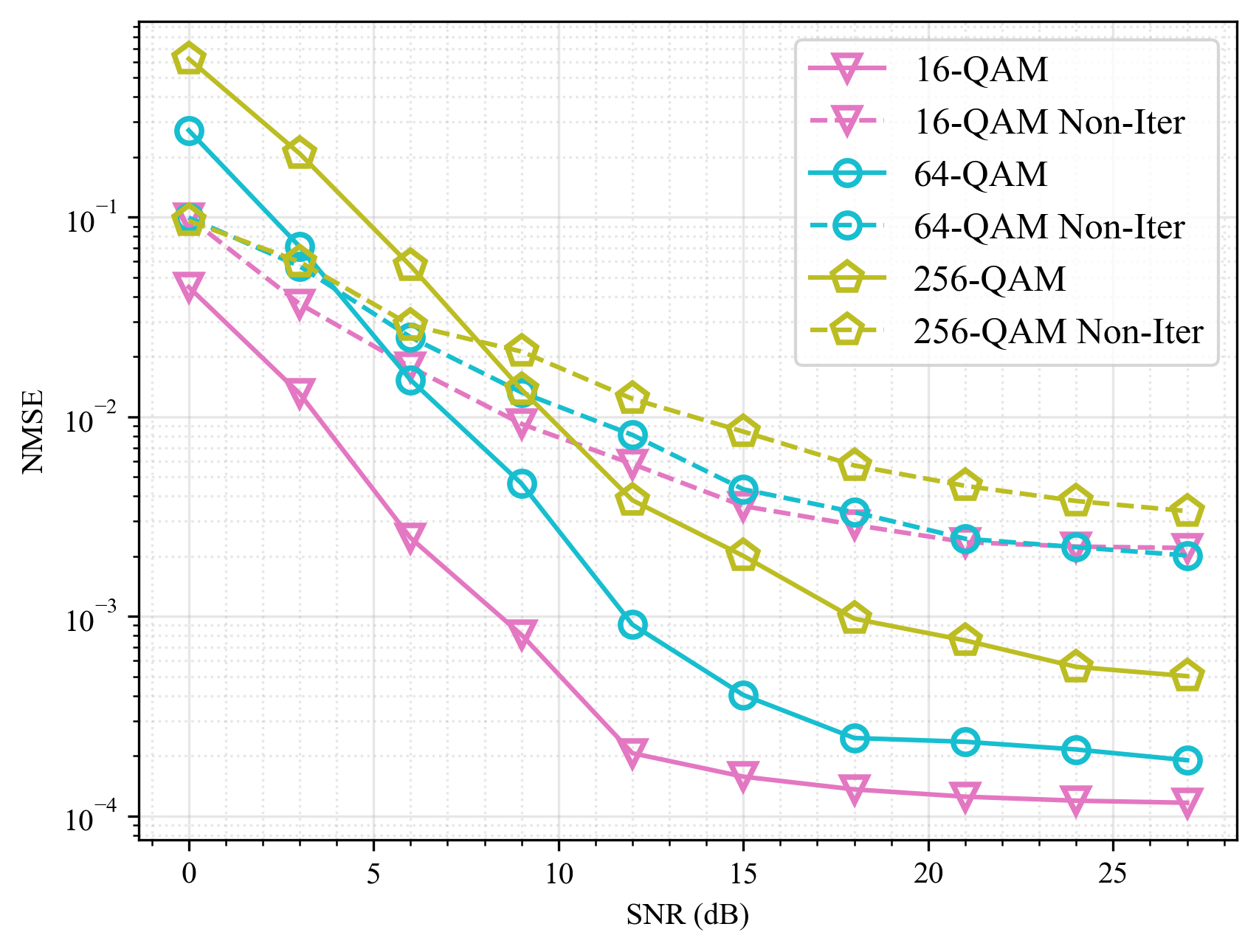}
    \centering
    \setlength{\abovecaptionskip}{0.cm}
    \caption{The NMSE of the proposed algorithm with and without iterative JCESD.} 
    \label{fig12}
\end{figure}

The proposed algorithm comprises two main components. 
First, semi-blind JCESD are performed based on the affine invariance between the transmit and receive constellations.
The resulting estimates are then used as initial values in an iterative JCESD procedure to enhance accuracy.
Fig. \ref{fig12} illustrates the effect of the subsequent iterative process on CE accuracy in the 24-user scenario.
Lower modulation orders yield greater accuracy improvements through iteration. 
For high-order modulation schemes, iteration processing may degrade CE accuracy at low SNR levels. 
This behavior is attributed to the elevated SER of high-order modulation under low SNR, which leads to error propagation during iterations and a corresponding decline in CE accuracy. 
Nevertheless, high-order modulation schemes are typically not employed in low-SNR environments.
At high SNR levels, the proposed iterative method achieves a reduction in NMSE of approximately 10–15 dB.
\par

\begin{figure}[ht]
    \centering
    \setlength{\abovecaptionskip}{0.cm}
    \includegraphics[width=0.48\textwidth]{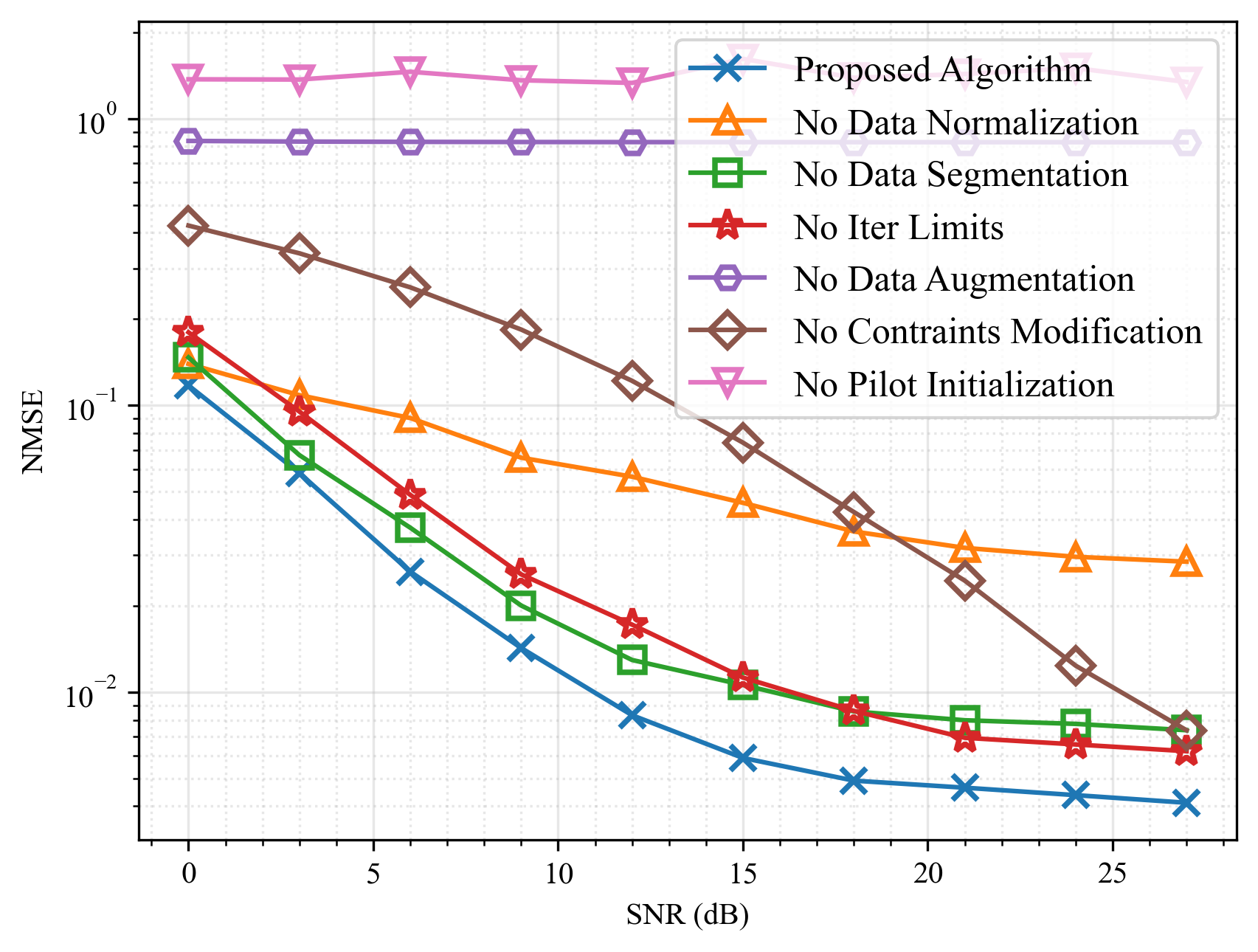}
    \setlength{\abovecaptionskip}{0.cm}
    \caption{NMSE performance comparison of the proposed method without iterative JCESD and its four simplified variants. There is 24 users and the MCS index is 15 (64-QAM).}
    \label{fig13}
\end{figure}

To further assess the contribution of each algorithmic component, an ablation study was conducted, as shown in Fig. \ref{fig13}. Specifically, removing the data normalization and pilot-based initialization modules lead to a marked degradation in CE accuracy, this highlights the critical role of scaling and initialization in facilitating convergence. The absence of data segmentation results in diminished performance due to insufficient granularity in addressing frequency selectivity. Meanwhile, reducing the degrees of freedom in the iterations leads to premature convergence, preventing full exploitation of the iterative refinement. Notably, excluding the data augmentation mechanism yields the most significant performance loss, underscoring its role in enhancing estimation robustness under sample-scarce conditions. Furthermore, replacing the proposed constraint design with the conventional constraint used in \cite{ref1} (denoted as "Original Constraints") leads to a considerable NMSE increase across all SNRs.
These findings collectively affirm the synergistic interplay among the algorithmic modules and highlight their necessity for achieving optimal CE performance.

\section{Conclusion}
\label{sec5}
In this paper, a semi-blind JCESD algorithm based on the affine invariance of \textit{M}-QAM constellation was proposed for MU massive MIMO downlink systems with large user populations. 
We formulated JCESD as a constellation fitting problem, and futher designed an optimization program tailored for high-order modulations by leveraging the geometric structure of QAM constellation.
The proposed aimed to reduce pilot overhead while maintaining the accuracy of CESD, thereby enhancing the throughput of the communication system.
Compared to existing blind and semi-blind methods, the proposed algorithm was extended to support high-order modulation schemes.
Numerical results demonstrated the proposed algorithm achieved significant throughput gains over pilot-based methods commonly employed in current communication standards, under various SNR conditions and system configurations.


%

\appendices
\section{Effect of Pilot-Based Initialization}
\label{appendixA}

According to (\ref{e42}), the equivalent channel estimate \(\mathbf{\hat H}_{\text{eq}}\) is obtained by the proposed method through LS estimation utilizing \(N_\text{s}\) pilot symbols. The initial iterate for the optimization \(\mathbf{U}_0={\left( {{{\mathbf{\hat H}}_{{\text{eq}}}}} \right)^{ - 1}}\) and denote the global optimum by \(\mathbf{U}^*={{\mathbf{H}}_\text{eq}^{ - 1}}\). Compared to a random initialization, pilot-based initialization yields \(\mathbf{U}_0\) systematically closer to \(\mathbf{U}^*\), thereby significantly increasing the probability that the SQP solver converges to a stationary point near \(\mathbf{U}^*\) \cite{ref40}. \par
For \(\Delta  = {{{\mathbf{\hat H}}}_{{\text{eq}}}} - {{\mathbf{H}}_{{\text{eq}}}}\) is sufficiently small, if \({\left\| {\mathbf{E}} \right\|_2}<1\) with \(\mathbf{E}={\mathbf{H}}_\text{eq}^{-1}\Delta\), utilizing Neumann series expansion we have
\begin{equation}
    \label{eA33}
    \setlength\abovedisplayskip{3pt}
    \setlength\belowdisplayskip{3pt}
    {\left( {{\mathbf{I}} + {\mathbf{E}}} \right)^{ - 1}} = \sum\limits_{m = 0}^\infty  {{{( - 1)}^m}{{\mathbf{E}}^m}}.
\end{equation}
\par
Truncating after the first-order term yields,
\begin{equation}
    \label{eA34}
    \setlength\abovedisplayskip{3pt}
    \setlength\belowdisplayskip{3pt}
    \begin{split}
        \mathbf{U}_0=&{\left( {{{\mathbf{H}}_{{\text{eq}}}} + \Delta } \right)^{ - 1}}\\
        =&{\left( {{\mathbf{I}} + {\mathbf{H}}_{{\text{eq}}}^{ - 1}\Delta } \right)^{ - 1}}{\mathbf{H}}_{{\text{eq}}}^{ - 1}\\
        =&{\mathbf{H}}_{{\text{eq}}}^{ - 1} - {\mathbf{H}}_{{\text{eq}}}^{ - 1}\Delta {\mathbf{H}}_{{\text{eq}}}^{ - 1} + O\left( {\left\| \Delta  \right\|_2^2} \right).
    \end{split}
\end{equation}
\par
Since \({{\mathbf{U}}^*} = {\mathbf{H}}_{{\text{eq}}}^{ - 1}\),
\begin{equation}
    \label{eA35}
    \setlength\abovedisplayskip{3pt}
    \setlength\belowdisplayskip{3pt}
    \mathbf{U}_0-\mathbf{U}^*=-\mathbf{U}^*\Delta\mathbf{U}^*+O\left({\left\| \Delta  \right\|_2^2}\right).
\end{equation}
\par
Using \({\left\| {{\mathbf{ABC}}} \right\|_\mathsf{F}} \leqslant {\left\| {\mathbf{A}} \right\|_2}{\left\| {\mathbf{B}} \right\|_\mathsf{F}}{\left\| {\mathbf{C}}\right\|_2}\), we obtain
\begin{equation}
    \label{eA36}
    \setlength\abovedisplayskip{3pt}
    \setlength\belowdisplayskip{3pt}
    {\left\| {{{\mathbf{U}}_0} - {{\mathbf{U}}^*}} \right\|_\mathsf{F}} \leqslant \left\| {{{\mathbf{U}}^*}} \right\|_2^2{\left\| \Delta  \right\|_\mathsf{F}} + O\left( {\left\| \Delta  \right\|_2^2} \right).
\end{equation}
\par
According to (\ref{e42}),
\begin{equation}
    \label{eA37}
    \setlength\abovedisplayskip{3pt}
    \setlength\belowdisplayskip{3pt}
    \begin{split}
        {{\mathbf{H}}^{{\text{eq}}}} + \Delta  =& {\left( {{\mathbf{P}}_{\text{t}}^\mathsf{H}{{\mathbf{P}}_{\text{t}}}} \right)^{ - 1}}{\mathbf{P}}_{\text{t}}^\mathsf{H}\left( {{{\mathbf{P}}_{\text{t}}}{{\mathbf{H}}^{{\text{eq}}}} + {\mathbf{N}}} \right)\\
        =&{\mathbf{H}}^{\text{eq}}+{\left( {{\mathbf{P}}_{\text{t}}^\mathsf{H}{{\mathbf{P}}_{\text{t}}}} \right)^{ - 1}}{\mathbf{P}}_{\text{t}}^\mathsf{H}\mathbf{N},
    \end{split}
\end{equation}
where \(\mathbf{N}\) is the noise matrix, which includes AWGN and IUI. Let the power of the pilot symbol be \(p\), then
\begin{equation}
    \label{eA38}
    \setlength\abovedisplayskip{3pt}
    \setlength\belowdisplayskip{3pt}
    \Delta={\left( {{\mathbf{P}}_{\text{t}}^\mathsf{H}{{\mathbf{P}}_{\text{t}}}} \right)^{ - 1}}{\mathbf{P}}_{\text{t}}^\mathsf{H}\mathbf{N}=\frac{1}{p}{\mathbf{P}}_{\text{t}}^\mathsf{H}\mathbf{N},
\end{equation}
\begin{equation}
    \label{eA39}
    \setlength\abovedisplayskip{3pt}
    \setlength\belowdisplayskip{3pt}
    {\left\| \Delta  \right\|_\mathsf{F}} \leqslant \sqrt {\frac{{{N_\text{s}}}}{p}} {\left\| {\mathbf{N}} \right\|_\mathsf{F}}.
\end{equation}
\par
Despite the non-convex nature of the optimization problem formulated in this paper, if the initial value \(\mathbf{U}_0\) lies within a neighborhood of radius \(r\) around \(\mathbf{U}^*\), the SLSQP iterations converge locally to \(\mathbf{U}^*\) \cite{ref40}. Pilot-based initialization ensures
\begin{equation}
    \label{eA391}
    \setlength\abovedisplayskip{3pt}
    \setlength\belowdisplayskip{3pt}
    {\left\| {{{\mathbf{U}}_0} - {{\mathbf{U}}^*}} \right\|_\mathsf{F}} \leqslant \sqrt {\frac{{{N_\text{s}}}}{p}} \left\| {{{\mathbf{U}}^*}} \right\|_2^2{\left\| {\mathbf{N}} \right\|_\mathsf{F}}+O\left( {\left\| \mathbf{N}  \right\|_2^2} \right).
\end{equation}
\par
The following analyzes the properties of an error \(\mathbf{R}=\mathbf{U}_0^\text{ran}-\mathbf{{U}^*}\) where \(\mathbf{U}_0^\text{ran}\) is a randomly generated initial matrix. 
\(\mathbf{U}_0^\text{ran}\) is drawn with each entry \({\left( {{\mathbf{U}}_0^{{\text{ran}}}} \right)_{ij}}\sim\mathcal{CN}(0,\alpha^2)\) i.i.d. where \(\alpha^2\) is the effective variance of the elements of \(\mathbf{U}_0^\text{ran}\). We have \(\mathbb{E}\left[ \mathbf{R}_{ij}  \right]= - \mathbf{U}^*_{ij}\) and \(\text{var}\left[ \mathbf{R}_{ij} \right]=\alpha^2\). 
Furthermore, \(2{{{\left| {{{\mathbf{R}}_{ij}}} \right|}^2}}/\alpha^2\) follows a scaled non-central chi-squared distribution (NC-\(\chi^2\)) with 2 degrees of freedom and non-centrality parameter \(\delta_{ij}\), where 
\begin{equation}
    \label{eA42}
    \setlength\abovedisplayskip{3pt}
    \setlength\belowdisplayskip{3pt}
    \delta_{ij}=\left( {\frac{{ - \Re \mathfrak{e}\left( \mathbf{U}_{ij}^* \right)}}{{\alpha /\sqrt 2 }}} \right)^2 + \left( {\frac{{ - \Im \mathfrak{m}\left( \mathbf{U}_{ij}^* \right)}}{{\alpha /\sqrt 2 }}} \right)^2 = \frac{{{{\left| \mathbf{U}^*_{ij} \right|}^2}}}{{{\alpha ^2}/2}}.
\end{equation}
\par
Now consider \(\left\|\mathbf{R}\right\|_\mathsf{F}^2={\sum\nolimits_{ij} {\left| {{{\mathbf{R}}_{ij}}} \right|} ^2}\), this is a sum of \(N_\text{s}^2\) independent random variables, each following a NC-\(\chi^2\) distribution. The sum follows a NC-\(\chi^2\) distribution with total degrees of freedom \(k_\text{totsl}=2N_\text{s}^2\) and non-centrality parameter \(\delta_{\text{total}}=2\left\|\mathbf{U}^*\right\|_\mathsf{F}^2/\alpha^2\). So, 
\begin{equation}
    \label{eA43}
    \setlength\abovedisplayskip{3pt}
    \setlength\belowdisplayskip{3pt}
    \frac{{\left\| {\mathbf{R}} \right\|_\mathsf{F}^2}}{{{\alpha ^2}/2}} \sim \chi_{2{N_\text{s}^2}}^{2}\left( {\frac{{\left\| {{{\mathbf{U}}^*}} \right\|_\mathsf{F}^2}}{{{\alpha ^2}/2}}} \right).
\end{equation}
\par
Let \(\epsilon^2\) be the square of the upper bound on the estimation error norm using pilot-based initialization, the probability \(p\left( {\left\| {\mathbf{R}} \right\|_\mathsf{F}^2 > {\epsilon^2}} \right)\) is the value of the cumulative distribution function (CDF) of the NC-\(\chi^2\) distribution \({F_{{\chi ^2}\left( {k,\delta } \right)}}\left( x \right)\) as
\begin{equation}
    \label{eA45}
    \setlength\abovedisplayskip{3pt}
    \setlength\belowdisplayskip{3pt}
    p\left( {\left\| {\mathbf{R}} \right\|_\mathsf{F}^2 > {\epsilon^2}} \right)={F_{{\chi ^2}\left( {2{N_\text{s}^2},\left\| {{{\mathbf{U}}^*}} \right\|_F^2/\left( {{\alpha ^2}/2} \right)} \right)}}\left( {\frac{{\epsilon^2}}{{{\alpha ^2}/2}}} \right).
\end{equation}
\par
The CDF can be expressed using the generalized Marcum Q-function \(Q_M(a, b)\). Specifically, \({F_{{\chi ^2}\left( {k,\delta } \right)}}\left( x \right)=1-Q_{k/2}\left(\sqrt{\delta}, \sqrt{x}\right)\). Thus,  
\begin{equation}
    \label{eA46}
    \setlength\abovedisplayskip{3pt}
    \setlength\belowdisplayskip{3pt}
    p={Q_{N_\text{s}^2}}\left( {\frac{{\sqrt 2 {{\left\| {{{\mathbf{U}}^*}} \right\|}_\mathsf{F}}}}{\alpha },\frac{\sqrt{2}\epsilon}{\alpha }} \right).
\end{equation}
\par
Marcum Q-function is monotonically increasing in \(a\) and decreasing in \(b\). This signifies that if the pilot-based initialization is highly accurate (i.e., small \(\epsilon\)) or \({{{\left\| {{{\mathbf{U}}^*}} \right\|}_\mathsf{F}}}\) is large, the probability that the error of the randomly generated initial value is higher than the error bound of the pilot-based initial value will significantly increase. Thus, starting at \(\mathbf{U}_0\) greatly increases the probability of convergence to \(\mathbf{U^*}\) under a non-convex condition.





\ifCLASSOPTIONcaptionsoff
  \newpage
\fi


\begin{thebibliography}{99}

\bibitem{ref13}
X. Lin, S. Wu, C. Jiang, L. Kuang, J. Yan and L. Hanzo, ``Estimation of broadband multiuser millimeter wave massive MIMO-OFDM channels by exploiting their sparse structure,'' \textit{IEEE Trans. Wireless Commun.}, vol. 17, no. 6, pp. 3959-3973, Jun. 2018.

\bibitem{ref14}
W. Jiang and H. D. Schotten, ``Cell-free massive MIMO-OFDM transmission over frequency-selective fading channels,'' \textit{IEEE Commun. Lett.}, vol. 25, no. 8, pp. 2718-2722, Aug. 2021.

\bibitem{ref16}
S. Wu, L. Kuang, Z. Ni, J. Lu, D. Huang and Q. Guo, ``Low-complexity iterative detection for large-scale multiuser MIMO-OFDM systems using approximate message passing,'' \textit{IEEE J. Sel. Topics Signal Process}, vol. 8, no. 5, pp. 902-915, Oct. 2014.

\bibitem{ref36}
T. Iwakuni, K. Maruta, A. Ohta, Y. Shirato, T. Arai and M. Iizuka, ``Inter-user interference suppression in time varying channel with null-space expansion for multiuser massive MIMO,'' \textit{2015 IEEE 26th Annual International Symposium on Personal, Indoor, and Mobile Radio Communications (PIMRC)}, Hong Kong, China, 2015, pp. 558-562.

\bibitem{ref37}
H. Q. Ngo, E. G. Larsson and T. L. Marzetta, ``Energy and spectral efficiency of very large multiuser MIMO systems,'' \textit{IEEE Trans. Commun.}, vol. 61, no. 4, pp. 1436-1449, Apr. 2013.

\bibitem{ref38}
J. -C. Shen, J. Zhang and K. B. Letaief, ``Downlink user capacity of massive MIMO under pilot contamination,'' \textit{IEEE Trans. Wireless Commun.},  vol. 14, no. 6, pp. 3183-3193, Jun. 2015.

\bibitem{ref19}
3GPP TS 38.211, ``NR; physical channels and modulation (release 18),'' 3rd Generation Partnership Project; Technical Specification Group Radio Access Network.

\bibitem{ref20}
Y. Liu and F. Wang, ``Blind channel estimation and data detection with unknown modulation and coding scheme,'' \textit{IEEE Trans. Commun.}, vol. 72, no. 5, pp. 2595-2609, May 2024.

\bibitem{ref21}
A. A. Siddig, \textit{et al.}, ``Efficient blind channel estimation for IRS-assisted joint communication and sensing in 6G networks,'' \textit{2024 IEEE Middle East Conference on Communications and Networking (MECOM)}, Abu Dhabi, United Arab Emirates, 2024, pp. 309-314.

\bibitem{ref10}
L. Tong, R. . -w. Liu, V. C. Soon and Y. . -F. Huang, ``Indeterminacy and identifiability of blind identification,'' \textit{IEEE Trans. Circuits Syst. I, Reg. Papers1}, vol. 38, no. 5, pp. 499-509, May 1991.

\bibitem{ref22}
J.-L. Yu and W.-R. Kuo, ``Fast semi-blind channel estimation for MIMO-OFDM systems with virtual carriers,'' \textit{2012 1st IEEE International Conference on Communications in China (ICCC)}, Beijing China, 2012, pp. 356-361.

\bibitem{ref23}
K. N. Aliyu, A. Lawal, K. Abed-Meraim and A. Zerguine, ``Fast subspace-based semi-blind channel estimation for MIMO-OFDM communications,'' \textit{2023 31st European Signal Processing Conference (EUSIPCO)}, Helsinki, Finland, 2023, pp. 1460-1463.

\bibitem{ref24}
Y. Li, Y. Jiang, X. Zhu, S. Sun and V. K. N. Lau, ``Semi-blind channel estimation for DCO-OFDM VLC systems,'' \textit{ICC 2024 - IEEE International Conference on Communications}, Denver, CO, USA, 2024, pp. 3707-3712.

\bibitem{ref25}
P. Singh, S. Srivastava, A. K. Jagannatham and L. Hanzo, ``Second-order statistics-based semi-blind techniques for channel estimation in millimeter-wave MIMO analog and hybrid beamforming,'' \textit{IEEE Trans. Commun.}, vol. 68, no. 11, pp. 6886-6901, Nov. 2020.

\bibitem{ref26}
S. -H. Fang, J. -Y. Chen, J. -S. Lin, M. -D. Shieh and J. -Y. Hsu, ``Blind channel estimation for CP/CP-free OFDM systems using subspace approach,'' \textit{2015 IEEE 81st Vehicular Technology Conference (VTC Spring)}, Glasgow, UK, 2015, pp. 1-5.

\bibitem{ref27}
S. Gulomjon, F. Yongqing, J. Sangirov, F. Ye and A. Olmasov, ``A performance analysis of optimized semi-blind channel estimation method in OFDM systems,'' \textit{2017 19th International Conference on Advanced Communication Technology (ICACT)},  PyeongChang, Korea (South), 2017, pp. 907-912.

\bibitem{ref28}
Taejoon Kim and Iksoo Eo, ``Reliable blind channel estimation scheme based on cross-correlated cyclic prefix for OFDM system,'' \textit{2006 8th International Conference Advanced Communication Technology}, Phoenix Park, Korea (South), 2006, pp. 3-5.

\bibitem{ref29}
O. Rekik, K. N. Aliyu, B. M. Tuan, K. Abed-Meraim and N. L. Trung, ``Fast subspace-based blind and semi-blind channel estimation for MIMO-OFDM Systems,'' \textit{IEEE Trans. Wireless Commun.},  vol. 23, no. 8, pp. 10247-10257, Aug. 2024.

\bibitem{ref30}
Zhang B, Yu J L and Kuo W, ``Fast convergence on blind and semi-blind channel estimation for MIMO–OFDM systems,'' \textit{Circuits, Systems, and Signal Processing}, vol. 34, pp. 1993-2013, Jun. 2015.

\bibitem{ref1}
T. R. Dean, M. Wootters and A. J. Goldsmith, ``Blind joint MIMO channel estimation and decoding,'' \textit{IEEE Trans. Inf. Theory}, vol. 65, no. 4, pp. 2507-2524, Apr. 2019.

\bibitem{ref31}
T. R. Dean, J. R. Perlstein, M. Wootters and A. J. Goldsmith, ``Fast blind MIMO decoding through vertex hopping,'' \textit{IEEE Trans. Wireless Commun.}, vol. 18, no. 7, pp. 3669-3682, Jul. 2019.

\bibitem{ref39}
E. Nayebi and B. D. Rao, ``Semi-blind channel estimation for multiuser massive MIMO systems,'' \textit{IEEE Trans. Signal Process.}, vol. 66, no. 2, pp. 540-553, Jan. 2018.

\bibitem{ref32}
R. W. Heath, N. González-Prelcic, S. Rangan, W. Roh and A. M. Sayeed, ``An overview of signal processing techniques for millimeter wave MIMO systems,'' \textit{IEEE J. Sel. Topics Signal Process.}, vol. 10, no. 3, pp. 436–453, Apr. 2016.

\bibitem{ref33}
L. Yan, C. Han, and J. Yuan, ``A dynamic array-of-subarrays architecture and hybrid precoding algorithms for terahertz wireless communications,'' \textit{IEEE J. Sel. Areas Commun.}, vol. 38, no. 9, pp. 2041–2056, Sep. 2020.

\bibitem{ref34}
F. Sohrabi and W. Yu, ``Hybrid digital and analog beamforming design for large-scale antenna arrays,'' \textit{IEEE J. Sel. Topics Signal Process.}, vol. 10, no. 3, pp. 501-513, Apr. 2016.

\bibitem{ref3}
L. Sun and M. R. McKay, ``Eigen-based transceivers for the MIMO broadcast channel with semi-orthogonal user selection,'' \textit{IEEE Trans. Signal Process.}, vol. 58, no. 10, pp. 5246-5261, Oct. 2010.

\bibitem{ref2}
H. Li, S. Wu, S. Zheng, C. Jiang, H. Yang and W. Zhang, ``Dual-driven learning for channel estimation of massive MIMO systems with one-bit ADCs,'' \textit{IEEE Wireless Commun. Lett.}, vol. 13, no. 4, pp. 1108-1112, Apr. 2024.

\bibitem{ref5}
T. A. Abose, T. O. Olwal, M. R. Hassen and E. S. Bekele, ``Performance analysis and comparisons of hybrid precoding scheme for multi-user mmwave massive MIMO system,'' \textit{2022 3rd International Conference for Emerging Technology (INCET)}, Belgaum, India, 2022, pp. 1-6.

\bibitem{ref4}
J. Rodríguez-Fernández, N. González-Prelcic, K. Venugopal and R. W. Heath, ``Frequency-domain compressive channel estimation for frequency-selective hybrid millimeter wave MIMO systems,'' \textit{IEEE Trans. Wireless Commun.}, vol. 17, no. 5, pp. 2946-2960, May 2018.

\bibitem{ref6}
W. Tan, S. D. Assimonis, M. Matthaiou, Y. Han, X. Li and S. Jin, ``Analysis of different planar antenna arrays for mmwave massive MIMO systems,'' \textit{2017 IEEE 85th Vehicular Technology Conference (VTC Spring)}, Sydney, NSW, Australia, 2017, pp. 1-5.

\bibitem{ref9}
S. Talwar, M. Viberg and A. Paulraj, ``Blind separation of synchronous co-channel digital signals using an antenna array. I. Algorithms,'' \textit{IEEE Trans. Signal Process.}, vol. 44, no. 5, pp. 1184-1197, May 1996.

\bibitem{ref7}
D. Kraft, ``A software package for sequential quadratic programming,'' \textit{Deutsche Forschungs- und Versuchsanstalt für Luft- und Raumfahrt}, Report DFVLR-FR 88-28, 1988.

\bibitem{ref8}
R. E. Perez, P. W. Jansen and J. R. R. A. Martins, ``pyOpt: A Python-based object-oriented framework for nonlinear constrained optimization,'' \textit{Structures Multidiscipl. Optim.}, vol. 45, no. 1, pp. 101–118, 2012.

\bibitem{ref35}
3GPP TS 38.214, ``NR; physical layer procedures for data (release 18),'' 3rd Generation Partnership Project; Technical Specification Group Radio Access Network.

\bibitem{ref11}
J. Hoydis \textit{et al.} ``Sionna: an open-source library for next-generation physical layer research'', 2022, \textit{arXiv: 2203.11854}.

\bibitem{ref12}
J. Beek, O. Edfors, M. Sandell, S. K. Wilson and P. o. Borjesson, ``On channel estimation in OFDM systems'', \textit{Proc. of IEEE 45th Vehicular Technology Conference}, pp. 815-819, July 1995.

\bibitem{ref40}
J. Nocedal and S. J. Wright, \textit{Numerical Optimization}, New York, NY, USA: Springer, 2006.


\end{thebibliography}
\end{document}